\pdfoutput=1     
\documentclass[twocolumn]{aastex7}

\usepackage{graphicx}
\usepackage{amsmath}
\usepackage{booktabs}

\newcommand{\Tpot}{T_{\mathrm{pot}}}

\newcommand{\Tcrit}{T_{\mathrm{crit}}}

\newcommand{\xsat}{x_{\mathrm{sat}}}
\newcommand{\Fint}{F_{\mathrm{int}}}

\newcommand{\Msil}{M_{\mathrm{sil}}}

\newcommand{\Le}{\mathrm{Le}}

\newcommand{\kappath}{\kappa_{\mathrm{th}}}
\newcommand{\Dhyd}{D_{\mathrm H_2}}
\newcommand{\km}{k_m}

\newcommand{\dd}{\mathrm{d}}
\newcommand{\Sc}{\mathrm{Sc}}

\newcommand{\kex}{k_{\mathrm{L}}}
\newcommand{\vconv}{v_{\mathrm{conv}}}

\newcommand{\dth}{\delta_\theta}
\newcommand{\dc}{\delta_c}

\shorttitle{}

\shortauthors{Young}

\begin{document}

\title{%
  \fontsize{15}{20}\selectfont
  Hydrogen Engulfment into Sub-Neptune Cores through Magma Ocean Convective Surface Renewal%
  \par
}


\author[0000-0002-1299-0801]{Edward D. Young}
\affiliation{Department of Earth, Planetary, and Space Sciences, University of California, Los Angeles, CA 90095, USA}
\email[show]{eyoung@epss.ucla.edu}
\correspondingauthor{Edward D. Young}

\begin{abstract}
A convective surface renewal explanation for ingassing of hydrogen from H$_2$-rich envelopes into molten cores of growing planets is presented, with particular application to sub-Neptunes. The ingassing of hydrogen occurs at a moving front comprising the upper boundary of the magma ocean that engulfs hydrogen from the envelope above as it moves.  The mechanism of engulfment is surface renewal as the upper diffusive boundary layer is diluted by convection. Surface renewal results in ingassing even where the stabilizing buoyancy of the low-density hydrogen neutralizes the destabilizing thermally-driven buoyancy. It is ultimately driven by cooling.  Beginning during the gaseous disk phase, when hydrogen is effectively available in infinite supply, sub-Neptunes can acquire enough hydrogen to permit global core-envelope equilibrium. The structures are therefore consistent with hydrogen-rich, supercritical magma oceans overlain by hydrogen-rich envelopes. The process of surface renewal has implications for other magma ocean - atmosphere interactions in general. 
\end{abstract}


\section{Introduction}

Interest in reactions between melts and hydrogen in a planetary context dates back decades \citep{Sasaki_1990, Ikoma2006}. The prevalence of gas-dwarf sub-Neptunes among planets in the Galaxy has provided new impetus for  studies of reactions between silicate-rich magma oceans and overlying hydrogen-rich envelopes \citep{Kite2019, Kite_2020b, Schlichting_Young_2022, Young_Nature_2023, Young_2024, Miozzi2025}.  Recent studies show that there are strong chemical potential driving forces for infusing molten cores with hydrogen \citep[e.g.,][]{Kite2019}. Most recently, advances in our understanding of the miscibility of hydrogen, silicates, and iron at relevant conditions have led to the inference that the boundary between magma oceans and overlying H$_2$-rich envelopes may be defined by the temperature and pressure-dependent phase change (binodal) between supercritical mixtures of silicate, iron and hydrogen, and exsolved, distinct silicate-rich and hydrogen-rich phases \citep{Young_2024, Young2025_Differentiation, gilmore_core-envelope_2025, Rogers_2025_redefine,Soutenburg_2026}.

These studies tacitly assume that planets can form with abundant hydrogen in their cores, where here  the term ``core" refers to the condensed molten rock-like phase or phases that lie beneath dense hydrogen-rich envelopes.  The envelopes limit the cooling of the gas-dwarf planets, permitting magma oceans to persist for billions of years \citep[e.g.,][]{ginzburg2016a}, depending on conditions. The physical processes leading to this intermingling of hydrogen and the silicate-rich cores have not been established, however. 

One possible mechanism for introducing hydrogen into cores is giant impacts between protoplanets with hydrogen atmospheres during planet formation. Simulations suggest that energetic collisions force materials to supercritical states where liquid and vapor merge, leading to intimate mixing of volatile components with refractory materials associated with condensed cores \citep{Lock2018, CaracasStewart2023, Roche2025}. While viable, the details of collision-induced mixing remain somewhat vague, in part because SPH simulations are not ideal for resolving mixing among materials with high density contrasts. 

More passive diffusive ingassing of hydrogen into melts has been considered.  A possible barrier to the efficacy of this process is the formation of a low-density upper boundary layer in the melt due to the lowering of melt densities by incorporation of hydrogen. \citet{Modirrousta-Galian_2025} evaluated whether a low-density surface layer could inhibit hydrogen ingassing into a terrestrial magma ocean. In their model, atmospheric H$_2$ reacts with FeO in the surface melt to form H$_2$O and metallic iron; rapid segregation of the metal leaves an FeO-depleted, H$_2$O-bearing layer that is less dense than the underlying magma. They assessed its stability using thermochemical Rayleigh-number, oscillatory double-diffusive, and convective-shear criteria, concluding that plausible thermal contrasts are insufficient to disrupt the buoyant layer. The resulting stagnant layer restricts continued reaction to the limited FeO inventory of the thin interfacial region, greatly reducing water production relative to a fully mixed magma ocean. Their calculation therefore represents the limit where the low-density boundary layer is entirely stagnant. 

The stagnant fixed boundary layer need not be the case, however.  The limits for ingress of hydrogen imposed by a stagnant low-density layer at the top of the magma oceans are diminished if surface renewal is considered. Surface renewal refers to the dilution of the low-density melt as the convecting interior periodically sweeps the near-surface layers away, replacing them with fresh, well-mixed parcels of melt.  Diffusion remains the mechanism for transport across the boundary, but advection supplies fresh material simultaneously, decreasing the density contrast and maintaining the chemical potential driving force for diffusion.  Advection determines how long the fluid resides at the surface. The concept of surface renewal has been applied in this context previously.  In these earlier models, surface renewal was attributed to impacts \citep{Olson_2018} or wind-based air--sea gas-transfer scalings \citep{Olson_2019}. 

Here, we explore a Danckwerts-type surface-renewal model \citep{Danckwerts_1951} that couples magma renewal to convective overturn, compositional stabilization, nonideal hydrogen diffusion, and evolution of the H$_2$--silicate phase boundary. The calculation invokes no exogenous resurfacing mechanism, such as impacts. Instead, renewal is driven by the magma ocean's own convection. It therefore tests whether substantial hydrogen ingassing is an intrinsic part of planetary growth, operating over the several-million-year lifetime of the protoplanetary disk and potentially continuing for tens of millions of years thereafter.

The Danckwerts surface renewal (or penetration) concept replaces fixed-boundary layer models with a sequence of brief contacts, whereby convection brings fresh parcels of melt to the surface. It is a natural consequence of an actively convecting system. Near the surface, exchange by diffusion between parcels of melt and the overlying envelope occurs for the brief period that the melt resides adjacent the thin diffusive boundary layer, whereupon convection removes the parcels. Rather than a single stagnant boundary layer, the boundary is continuously replenished, preventing stalling of the process that might otherwise arise by elimination of the chemical potential driving force for diffusion.  Simultaneously, convection conveys the low-density component acquired by diffusion downward into the bulk melt. In effect the single contact time is replaced by a statistical renewal where parcels are replaced at random so their ages with respect to exposure to the surface follow an exponential distribution.  

The surface renewal concept is applied here to the ingassing of hydrogen from the protoplanetary disk, through the hydrogen-rich envelope of the planet, and into the molten core below the envelope.  The process is shown to be a plausible mechanism for planets with molten cores to harbor large mass fractions of hydrogen without exogenous forcings such as collisions.  

The paper is organized as follows. Section 2 develops the concept of surface renewal, and Section 3 describes the hydrogen-engulfment front. Section 4 examines how thermal and compositional buoyancy compete to influence ingassing, while Section 5 considers the energetic constraints on mixing in the magma-ocean phase. Section 6 presents the model calculations, followed by the results in Section 7 and their discussion in Section 8. Section 9 summarizes the conclusions.

\section{Surface Renewal Process}
\label{sec:surface_renewal}
The crucial parameter for the surface renewal process is the mass transfer coefficient, $\kex$.  The classical two-layer (or two-film) theory pictures a fixed
stagnant thin layer, or film, of thickness $\delta$ crossed by steady diffusion, resulting in
$\km\propto\Dhyd$ \citep{Whitman_1923} where we are concerned in this context with the diffusion of H$_2$ across an upper boundary layer of a melt. \cite{Higbie1935} discarded the steady film picture in favor of a single \emph{transient} contact of fixed duration $t_c$, predicting the weaker dependence on diffusivity, $\kex\propto\sqrt{\Dhyd}$. \cite{Danckwerts_1951} then replaced Higbie's one
fixed contact time with a \emph{distribution} of surface ages generated by random
renewal at a mean rate $s$, yielding the form this model adopts, $\kex=\sqrt{\Dhyd\,s}$. The fundamental divide in treatments is between Whitman and the two renewal models---the linear $\Dhyd$ of film theory versus the $\sqrt{\Dhyd}$ of penetration and surface renewal. These differences are described in greater detail below.

The renewal picture is explained here in the context of engulfment of H$_2$ into an initially hydrogen-poor molten core. Consider a parcel of well-mixed interior melt, initially nearly
hydrogen-free, that is brought into contact with the melt-envelope boundary for a finite contact
time $t_c$. During this interval, transport normal to the thin boundary is assumed to be
controlled by molecular diffusion. Let $z$ denote distance measured downward from the
interface, with $z=0$ at the boundary. The concentration then satisfies Fick's second law,
\begin{equation}
\frac{\partial C}{\partial t} = D\,\frac{\partial^2 C}{\partial z^2},
\end{equation}
where $D$ is the molecular diffusivity of hydrogen in the melt and $C=\rho X$ is the
hydrogen concentration expressed as a partial mass density ($X$ the mass fraction). Here,  concentration is used as a surrogate for chemical potential, but we will return to the distinction later. The parcel arrives
carrying the well-mixed interior value $C_{\mathrm{melt}}=\rho X_{\mathrm{melt}}$, which
for an initially dry interior is close to zero,
\begin{equation}
C(z,0)=C_{\mathrm{melt}},
\end{equation}
while the interface is held at the \emph{hydrogen-rich} surface value
$C_{\mathrm{surf}}=\rho X_{\mathrm{surf}}$ imposed by the overlying envelope.  

The fact that hydrogen and silicate (and Fe) are fully miscible at temperatures and pressures above the binodal prescribed by ab initio molecular dynamics calculations is used here \citep{gilmore_core-envelope_2025}, as envisioned by \cite{Young_2024}. When the melt phase is at pressures and temperatures greater than the composition-dependent critical point, the thermodynamic driving force is for the hydrogen-silicate ($\pm$ Fe) boundary to be fully miscible. Therefore, under these conditions, the surface composition equals the bulk system composition, $X_{\rm surf} = X_{\mathrm{bulk}}$, where the mixture is supercritical/miscible.  The assumption here is that the melt phase and envelope phase influencing the interface between them are in the main well mixed by convection (a bias in one or the other does not substantially change the picture described here). Once temperatures and pressures dip below the binodal for the two phases, 
the melt-side binodal concentration, $x_{\mathrm{sat}}$, determines the interface composition, $X_{\rm surf} = x_{\rm sat}$, such that 
\begin{equation}
C(0,t)=C_{\mathrm{surf}}\;>\;C_{\mathrm{melt}} .
\end{equation}
Far from the interface, the melt is well mixed, and 
\begin{equation}
C(z\rightarrow\infty,t)=C_{\mathrm{melt}} .
\end{equation}

The difference in surface concentration and well-mixed melt concentration is afforded by the diffusion across the boundary. Two different solutions illustrate the distinction between a stagnant boundary layer and one experiencing surface renewal, and thus the distinct mass transfer coefficients. The issue is, does a patch of surface melt sit in contact long enough for the profile
to reach \emph{steady state}, or is it swept away while the profile is still \emph{transient}?
The two limits carry different powers of $D$, and that difference is the essence of
surface-renewal theory.

\subsection{Steady Boundary Layer, the Steady-state Case}
If the surface layer were a fixed, persistent
boundary, or ``film", of thickness $\delta$ across which diffusion has come to steady state, then
$\partial C/\partial t=0$ and Fick's second law reduces to $\partial^2C/\partial z^2=0$.
Integrating twice, the concentration profile is linear between the interface value at
$z=0$ and the interior value at $z=\delta$,
\begin{equation}
C(z) = C_{\mathrm{surf}}-\left(C_{\mathrm{surf}}-C_{\mathrm{melt}}\right)\frac{z}{\delta}.
\end{equation}
Differentiating, the gradient is the same at every depth, a constant fixed entirely by the
film thickness $\delta$,
\begin{equation}
\left.\frac{\partial C}{\partial z}\right|_{z=0}
= -\,\frac{C_{\mathrm{surf}}-C_{\mathrm{melt}}}{\delta}.
\end{equation}
Fick's law then gives the interfacial flux, taken positive downward into the melt,
\begin{equation}
J = \frac{D\left(C_{\mathrm{surf}}-C_{\mathrm{melt}}\right)}{\delta},
\end{equation}
and defining the mass-transfer coefficient through
$J=k\,(C_{\mathrm{surf}}-C_{\mathrm{melt}})$ yields
\begin{equation}
k = \frac{D}{\delta}\ \propto\ D .
\end{equation}
The diffusivity enters \emph{linearly}, because the transport distance $\delta$ is independent of $D$.

\subsection{Renewal, the Transient Case}
Convection carries fresh parcels to the engulfment front, where they reside only briefly, and
sweeps them back down before any steady film can form. During that short contact a parcel
behaves as a \emph{semi-infinite} medium and the concentration follows the standard transient (penetration) solution,
\begin{equation}
C(z,t) = C_{\mathrm{surf}} - \left(C_{\mathrm{surf}}-C_{\mathrm{melt}}\right)\,
\operatorname{erf}\!\left(\frac{z}{2\sqrt{Dt}}\right)
\end{equation}
i.e., the textbook solution for diffusion into a
half-space held at a fixed surface concentration. The crucial feature is that the length
scale is not a fixed $\delta$ but the self-generated penetration depth
$\sqrt{Dt}$, the distance the front has diffused after a time $t$. Differentiating and
evaluating at the interface, the surface gradient therefore decays in time as the
front thickens,
\begin{equation}
\left.\frac{\partial C}{\partial z}\right|_{z=0}
= -\,\frac{C_{\mathrm{surf}}-C_{\mathrm{melt}}}{\sqrt{\pi D t}}.
\end{equation}
Fick's law, taken positive downward into the melt, then gives the instantaneous flux into the
parcel,
\begin{equation}
J(t) = \left(C_{\mathrm{surf}}-C_{\mathrm{melt}}\right)\sqrt{\frac{D}{\pi t}} .
\end{equation}
Defining the mass-transfer coefficient through
$J(t)=k(t)\,(C_{\mathrm{surf}}-C_{\mathrm{melt}})$ yields the penetration-theory result,
\begin{equation}
k(t) = \sqrt{\frac{D}{\pi t}}\ \propto\ \sqrt{D} .
\end{equation}
The two solutions of the same equation differ in what is held constant: the steady film
fixes a \emph{length}, the film thickness $\delta$, so $k=D/\delta\propto D$; the transient
contact instead fixes a \emph{time}, and lets the diffusion determine the length
$\sqrt{Dt}\propto\sqrt{D}$, so one power of $D$ is absorbed into the penetration depth and
$k\propto\sqrt{D}$. What fixes that time is the refresh rate as the melt is continually swept away and
replaced by fresh interior melt. The frequency of that replacement is the
\emph{surface-renewal rate}, $s$, the probability per unit time that a given surface element is
refreshed so that a parcel spends a characteristic contact time $t_c\sim 1/s$ at the interface
before being carried back down. It is this $s$, a consequence of the convective overturn, that supplies the
fixed timescale the transient solution relies on.  Next, the instantaneous coefficients for the distribution of contact times are averaged, yielding the surface-renewal result $k\propto\sqrt{D\,s}$.

\subsection{Danckwerts Surface Renewal Transfer Coefficient}
If every surface element is
exposed for the \emph{same} contact time $t_c$, evaluating the instantaneous
result at $t=t_c$ gives the Higbie (1935) transfer coefficient,
\begin{equation}
\,k_{\rm L}^{\mathrm{Higbie}} = \sqrt{\frac{D}{\pi t_c}}\,.
\end{equation}
Equivalently, the diffusive penetration depth after a contact time $t_c$ is of
order $\delta_D \sim \sqrt{\pi D t_c}$, so that the transfer coefficient may be
written as $k_{\rm L}\sim D/\delta_D\sim\sqrt{D/(\pi t_c)}$: it is prescribed by the
molecular diffusion length that develops during the finite exposure of a parcel,
not by the full thickness of the convective or thermal boundary layer. Averaging
the flux instead over the entire single exposure $0<t<t_c$,
\begin{equation}
\begin{split}
\overline{J} &= \frac{1}{t_c}\int_0^{t_c}\left(C_{\mathrm{surf}}-C_{\mathrm{melt}}\right)
\sqrt{\frac{D}{\pi t}}\,\mathrm{d}t \\
&= 2\left(C_{\mathrm{surf}}-C_{\mathrm{melt}}\right)\sqrt{\frac{D}{\pi t_c}},
\end{split}
\end{equation}
gives an average coefficient $\overline{k}_{\rm L}=2\sqrt{D/(\pi t_c)}$, larger by a
factor of two, the only difference between the instantaneous and exposure-
averaged Higbie estimates.

In a turbulently stirred fluid the surface elements are not all the same
age.  Rather, they are replaced at random where again $s$ is the
renewal rate, the probability per unit time that any surface element is swept away
and replaced by fresh, hydrogen-poor interior melt. With a constant $s$, such that $dN/dt = -s N$ where $N$ is the number of parcels of a cohort that are extant at the surface at age $t$, the fraction of surface with
exposure age between $t$ and $t+\dd t$ follows the exponential age distribution \citep{Danckwerts_1951}
\begin{equation}
\phi(t) = s\,e^{-st},
\qquad \int_0^\infty \phi(t)\,\dd t = 1 .
\end{equation}
The mean transfer coefficient is the instantaneous penetration coefficient 
$k(t)=\sqrt{D/\pi t}$ averaged over this distribution,
\begin{equation}
\begin{split}
\overline{k}_{\rm L}
   &= \int_0^\infty s\,e^{-st}\,\sqrt{\frac{D}{\pi t}}\;\dd t \\
   &= \sqrt{\frac{D}{\pi}}\;s\int_0^\infty t^{-1/2}e^{-st}\,\dd t \\
   &= \sqrt{\frac{D}{\pi}}\;s\,\sqrt\pi\,s^{-1/2}, 
\end{split}
\end{equation}
and with cancellations, we have the \emph{Danckwerts surface-renewal} coefficient
\begin{equation}
k_{\rm L} = \overline{k}_{\rm L} = \sqrt{D\,s}.
\end{equation}
This is the form of the transfer coefficient used here. The Schmidt number $\mathrm{Sc}=\nu/D$, the ratio of the melt's kinematic viscosity $\nu$ to the
hydrogen diffusivity $D$, measures how much
faster momentum diffuses than dissolved hydrogen, with $\mathrm{Sc}\gg1$ for a slow diffuser in a
viscous melt. Because $D=\nu/\mathrm{Sc}$, the Danckwerts coefficient may be written equivalently as
$\kex=\sqrt{D\,s}=\sqrt{\nu s/\mathrm{Sc}}\propto\mathrm{Sc}^{-1/2}$, the dimensionless form in which
the surface-renewal scaling is conventionally reported and empirically tested (a film law would give
$\mathrm{Sc}^{-1}$).  The $\sqrt{\Dhyd}$ (equivalently $\Sc^{-1/2}$) scaling afforded by the renewal picture has been
validated by gas--liquid absorption experiments \citep{Jahne1987, JahneHaussecker1998}
and, more recently, direct numerical simulation \citep{MAGNAUDET_CALMET_2006}.

\begin{figure*}[!t]
\centering
\includegraphics[width=0.8\textwidth]{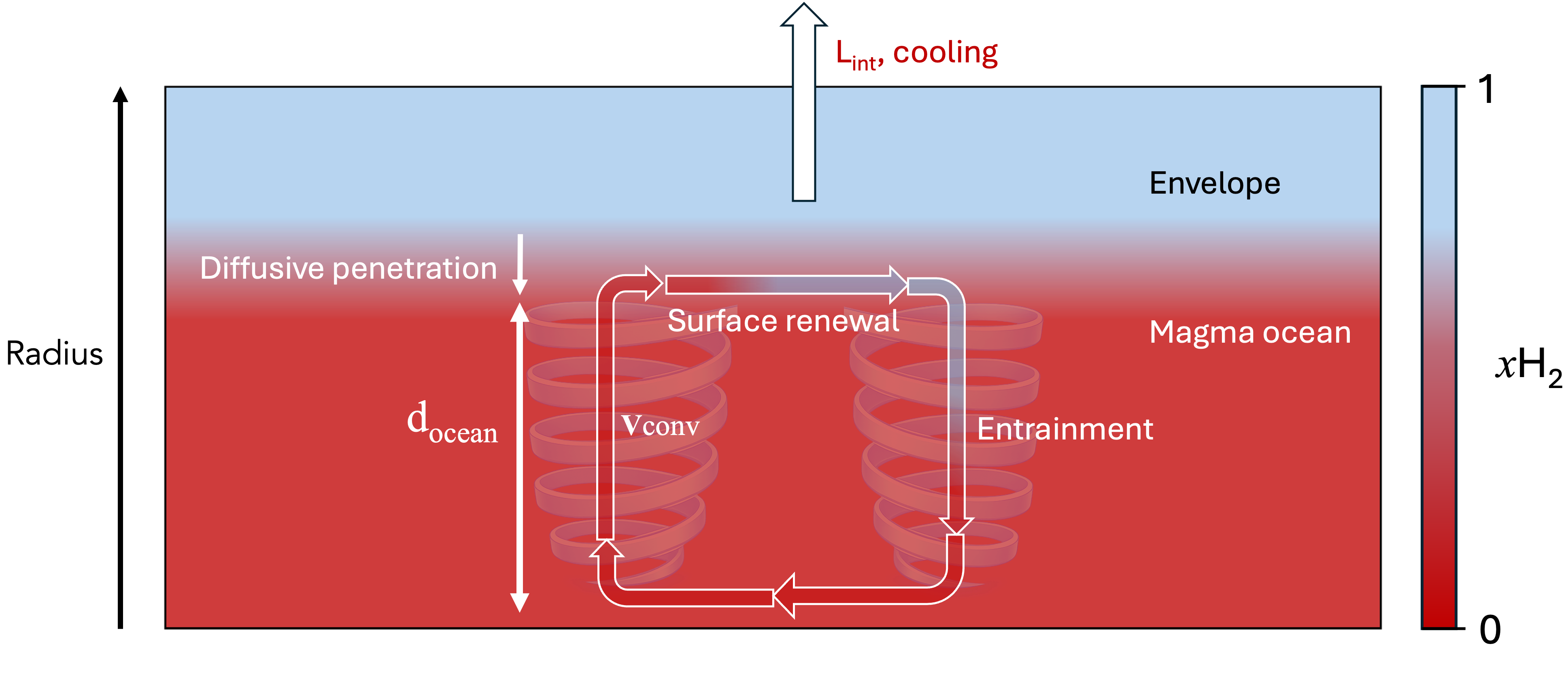}
\caption{Schematic illustrating convective surface renewal. Convective overturn entrains hydrogen delivered by diffusion from the envelope as the system cools. The spirals signify the effects of rotation on convection and the effect this has on the convective overturn velocity $v_{\rm conv}$ (see text). Color scheme illustrates relative concentrations of H$_2$ as mixtures of red and blue: red is ${\rm H}_2$-free melt, blue is pure ${\rm H}_2$ (+He in the case of the envelope). The magma ocean--envelope boundary is where the color switches abruptly to pure blue. }  
\label{fig:renewal}
\end{figure*}

The convective overturn frequency is adopted as the renewal rate,
\begin{equation}
s=\frac{\vconv}{L},\qquad L=d_{\mathrm{ocean}} .
\label{eq:danckwerts}
\end{equation}
Each renewal delivers a fresh,
\emph{hydrogen-poor} parcel of interior melt to the surface, which absorbs only a penetration load
of hydrogen during its brief contact $1/s$ before being swept back down
and replaced. The interior is thus fed hydrogen layer by layer (Figure \ref{fig:renewal}).

Predicting $s$ is  tantamount to predicting the scale of eddies that refresh the surface. The large-eddy closure \citep{Fortescue_1967} attributes the refresh rate to the energy-containing motions, $s\sim u_{\mathrm{rms}}/L_\infty$ with
$L_\infty$ the integral scale of the turbulence. For a magma ocean the energy-containing motions
are the buoyancy-driven cells and plumes, of aspect ratio near unity, whose integral scale
is the ocean depth and whose velocity is the convective overturn speed; identifying
$u_{\mathrm{rms}}$ with $\vconv$ and $L_\infty$ with $d_{\mathrm{ocean}}$ means  $s=\vconv/d_{\mathrm{ocean}}$ \citep{Solomatov2000, Solomatov_2015}.

A small-eddy alternative \citep{Lamont_1970}, developed for shear-driven interfaces,
would instead attribute the refresh rate to the smallest, Kolmogorov-scale, eddies and give a
larger mass-transfer coefficient. The relative efficacy of the two scales is evident with reference to the
convective Reynolds number $\mathrm{Re}=\vconv L/\nu$.  It can be shown that\footnote{The
large-eddy closure is $s\sim\vconv/L$; the small-eddy closure is
$s\sim1/\tau_\eta=(\varepsilon/\nu)^{1/2}$, the inverse Kolmogorov time, with dissipation
rate $\varepsilon\sim\vconv^{3}/L$. Their ratio is
$(\vconv^{3}/L\nu)^{1/2}/(\vconv/L)=(\vconv L/\nu)^{1/2}=\mathrm{Re}^{1/2}$, so that
$\kex\propto\sqrt{s}$ gives $\kex^{\rm eddy}/\kex^{\rm ocean}\sim\mathrm{Re}^{1/4}$.}
$\kex^{\mathrm{eddy}}/\kex^{\mathrm{ocean}}\sim\mathrm{Re}^{1/4}$. Nominal
magma-ocean values, including $L=d_{\rm ocean}\sim10^{6}$ to $10^{7}\,\mathrm{m}$, a water-like melt
kinematic viscosity $\nu\sim10^{-5}\,\mathrm{m^{2}\,s^{-1}}$ \citep{Karki_Stixrude_2010},
and an overturn speed $\vconv\sim0.1$ to $1\,\mathrm{m\,s^{-1}}$ \citep{Solomatov_2015}, give
$\mathrm{Re}\approx10^{10}$ to $10^{12}$, so the small-eddy closure for the mass transfer coefficient would exceed the
large-eddy one by $\kex^{\mathrm{eddy}}/\kex^{\mathrm{ocean}}\sim10^{2}$ to $10^{3}$. The larger transfer coefficient, however, is not what controls renewal for the magma ocean. 

We adopt the rate $s=\vconv/d_{\rm ocean}$ because a magma ocean is stirred
by coherent, depth-spanning convective cells rather than the 
shear cascade for which the small-eddy picture was developed, and whole-ocean renewal is
a consequence of the deep overturn that maintains the well-mixed interior, not the fastest
near-surface motions. Surface refresh by small eddies presumes each arriving parcel carries the well-mixed interior value $X_{\mathrm{melt}}$,
which holds only if the near-surface melt experiences the replenishment from depth that convective overturn provides. The delivery from depth by large-scale circulation to small eddies near the surface by necessity occurs in series, 
so their transfer coefficients add as a harmonic sum \citep{Danckwerts_1951},
\begin{equation}
\frac{1}{k_{\mathrm{L}}}=\frac{1}{k^{\mathrm{ocean}}_{\rm L}}+\frac{1}{k^{\mathrm{eddy}}_{\rm L}},
\label{eq:series}
\end{equation}
so the slower step controls the effective coefficient. Taking $L=d_{\mathrm{ocean}}$ identifies the renewal with
the slower overturn, collapsing the effective transfer coefficient to this scale.
Equating the renewal length with $d_{\mathrm{ocean}}$ is therefore not a statement about the size of the surface eddies but a manifestation of the depth over which the global overturn cycles bring fresh melt to the surface.

\section{Engulfment}
\label{sec:engulfment}
\subsection{Interfacial Flux}

The hydrogen transferred to the molten interior of the planet depends on the time available (e.g., disk lifetime), on how fast hydrogen is transferred across the boundary by diffusion and surface renewal, and on how fast the envelope-core interface advances.  The latter results from the addition of hydrogen to the melt phase when the conditions are supercritical with respect to melt-hydrogen mixing.  The melt consumes hydrogen in response to the chemical potential gradient across the melt-envelope interface, a manifestation of the thermodynamic drive to form the one stable phase out of the two unstable phases.  As the core amasses low-density hydrogen, it increases in mass and decreases in density.  This results in transforming the core-envelope interface into an engulfment front that consumes hydrogen as it moves.  We refer to the core-envelope interface as a ``front"  to emphasize that it is not an immutable surface, but rather evolves in composition and location with time. In the gaseous disk phase of planet accretion, hydrogen is in effect in infinite supply, and this front will move outward in the planet by consuming hydrogen at least as long as the disk is present; each parcel of hydrogen-rich envelope consumed by melt is replaced by accretion from the protoplanetary disk \citep{lee2015a,ginzburg2016a}. The rate of advancement of the engulfment front is described here. 

\begin{figure}[!t]
\centering
\includegraphics[width=0.9\columnwidth]{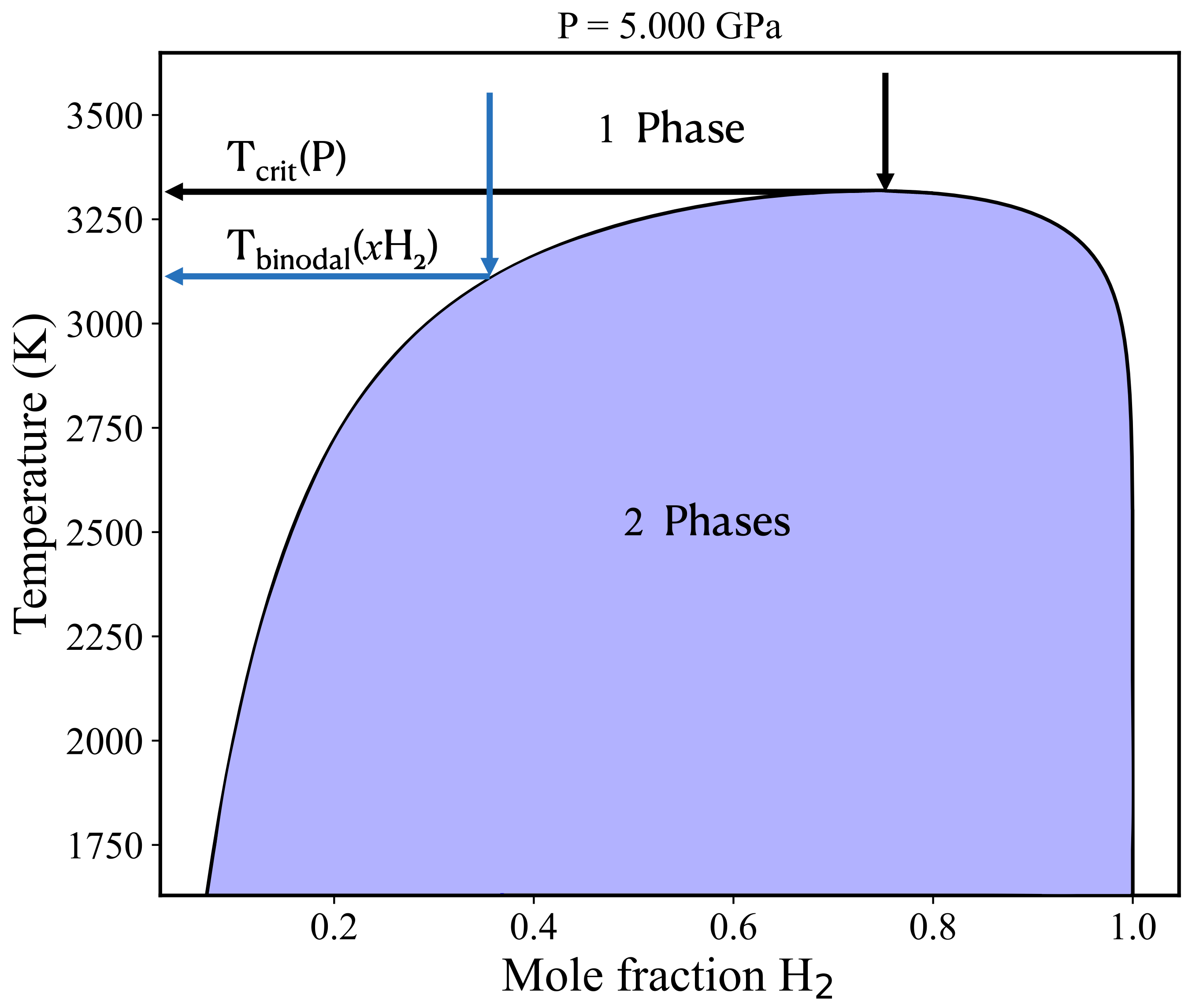}
\caption{Example phase diagram for the binary system MgSiO$_3$-H$_2$ illustrating the boundary between the one-phase, supercritical field, and the two-phase field where silicate and a distinct H$_2$-rich gas-like phase coexist stably. The boundary is the binodal. Two temperatures are used in this work to reference the binodal.  The consolute temperature, or critical temperature, $T_{\rm crit}$, is the highest temperature defining the crest of the binodal surface at a given pressure and is a useful reference because it is independent of composition.  The more general binodal temperature applies for a given composition along the binodal. Vertical arrows show cooling from the one-phase region to the two-phase region of the phase diagram.}    
\label{fig:phase_diagram}
\end{figure}

\begin{figure}[!t]
\centering
\includegraphics[width=0.99\columnwidth]{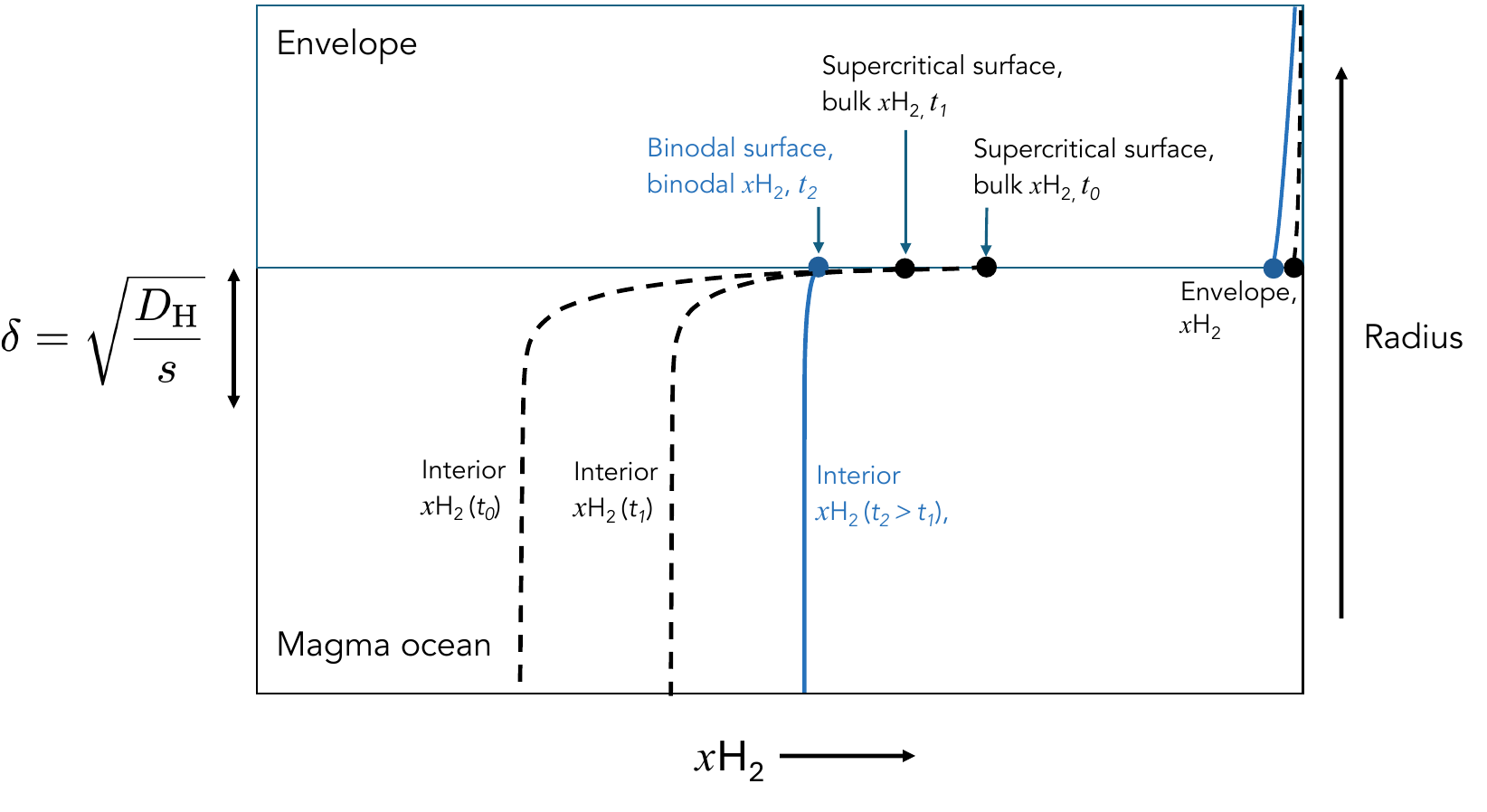}
\caption{Schematic illustrating the driving forces for ingassing of hydrogen by diffusion coupled with surface renewal. The abscissa is the concentration of hydrogen, and the ordinate is the radial distance from the center of the planet (not to scale). When the core is supercritical with respect to silicate-hydrogen mixing, the bulk hydrogen concentrations for the envelope and melt phases together define the target $x_{\rm H_2}$ at the melt-envelope interface. With time, e.g., in the sequence $t_0$, $t_1$, and $t_2$, the interior hydrogen concentration increases, the interface conditions eventually intersect the binodal, and the binodal melt $x_{\rm H_2}$ then defines the interface composition.  The width of the diffusive boundary layer is a function of the diffusivity and the renewal rate $s$.    }  
\label{fig:driving_forces}
\end{figure}

Hydrogen crosses the interface by surface-renewal mass transfer, as described above, and the flux is the product of a thermodynamic driving force and the transport that relieves it. Local equilibrium at the interface sets a target composition $X_{\rm surf}$, the hydrogen mass fraction the melt would have in equilibrium with the overlying envelope.  $X_{\rm surf}$ is the bulk hydrogen concentration defined by the well-mixed envelope and melt phases where the mixture is supercritical and there is a thermodynamic driving force to mix to a single phase, and it is the melt-side limb of the silicate-hydrogen binodal (the saturation solubility) where distinct silicate-rich and H$_2$-rich phases are thermodynamically stable. The well-mixed interior, at $X_{\rm melt}$, is
generally displaced from this value, and the departure $(X_{\rm surf}-X_{\rm melt})$ is the driving
force that is to first order the chemical-potential difference across the interface, linearized about
equilibrium (the Darken factor $\Phi$ embedded in the diffusivities $D$, restores its exact form, see below). The surface-renewal coefficient $\kex$ determines the rate at which fresh melt is carried toward that target, so the flux is the
product of a kinetic coefficient and a driving force,
\begin{equation}
J_{\rm H}=\rho\,\kex\,(X_{\rm surf}-X_{\rm melt}).
\label{eq:JH}
\end{equation}
Here $J_{\rm H}$ is the hydrogen mass flux
($\mathrm{kg\,m^{-2}\,s^{-1}}$), $\rho$ the melt density ($\mathrm{kg\,m^{-3}}$), $\kex$ the
surface-renewal mass-transfer coefficient, a velocity ($\mathrm{m\,s^{-1}}$), and
$(X_{\rm surf}-X_{\rm melt})$ the dimensionless H$_2$ mass-fraction difference between the interface and
the well-mixed interior. This equation applies to the thin compositional boundary layer separating the core and hydrogen-rich envelope of a growing gas dwarf planet. In the general ``hot start" case, this boundary is initially at temperatures and pressures greater than the consolute (or critical) temperature at a given pressure, where the mere presence of two phases ensures the system is out of physicochemical equilibrium (Figure \ref{fig:phase_diagram}). Later, the interface may cool to conditions consistent with the two-phase binodal. In both cases, this form of the driving force for chemical equilibrium holds, only the target $X_{\rm surf}$ differs (Figure \ref{fig:driving_forces}). The flux vanishes at equilibrium ($X_{\rm melt}=X_{\rm surf}$), is positive (ingassing) when the interior is undersaturated relative to the interface and negative (degassing) when oversaturated.  

Diffusion is driven by the chemical-potential gradient, so the chemical
interdiffusivity differs from the self-diffusivity by the Darken factor
\begin{equation}
  \Phi = 1 + \frac{\partial\ln\gamma_{\mathrm H_2}}{\partial\ln x}
  = \frac{x(1-x)}{RT}\,\frac{\partial^2 \hat{G}_{\mathrm{mix}}}{\partial x^2},
  \label{eq:darken}
\end{equation}
computed from the molar free energy of mixing, $\hat{G}_{\mathrm{mix}}$, that defines the binodal describing the miscibility of silicate and hydrogen, and evaluated at
$\xsat$ (just outside the spinodal, where $\Phi>0$). In the application here, for this strongly
asymmetric solution, $\Phi\sim2$--$3$ on the melt-side limb, an enhancement, but a marginal one. 
Nonetheless, in order to convert concentration gradients to chemical potential gradients, the chemical diffusivity is modified such that $D=\Phi\,D_{\rm self}$ where $D_{\rm self}$ is
the self-diffusivity of hydrogen.

With this hydrogen mass flux, the melt accumulates hydrogen according to
\begin{equation}
\frac{\dd X_{\rm melt}}{\dd t}=(1-X_{\rm melt})^{2}\,\frac{J_{\rm H}\,A_\varphi}{\Msil},
\label{eq:load}
\end{equation}
where $A_\varphi$ is the surface area at the interface.   

\subsection{The Engulfment Front: A Moving Phase Boundary}

As hydrogen dissolves and the melt thus acquires under-dense mass, the interface between the melt and envelope moves outward. The melt volume grows by the partial volume of the dissolved hydrogen, so global mass conservation
fixes the front velocity as the flux times the partial specific volume of dissolved hydrogen,
\begin{equation}
\dot r_\varphi=\bar v_{\rm H}\,J_{\rm H}\ .
\label{eq:rfdot}
\end{equation}
In principle, the partial specific volume follows from the melt equation of state,
\begin{equation}
\bar v_{\rm H}=\frac{1+(1-X_{\rm melt})\beta_X}{\rho},
\label{eq:vbar}
\end{equation}
where $\beta_X=-\rho^{-1}(\partial\rho/\partial X)$ is the fractional density reduction per unit dissolved
hydrogen. Dissolved H$_2$ has a large
volume effect. For example, at 6000 K and 3.5 GPa, the density of pure MgSiO$_3$ is 2500 kg m$^{-3}$ while the mixture with $4\%$ by mass H$_2$ has a density of 1350 kg m$^{-3}$ \citep{Young2025_Differentiation}. Therefore, with $\beta_X>0$ the front runs faster than $J_{\rm H}/\rho$ if $\rho$ were to be unaffected by the addition of hydrogen. In practice, because the engulfment-front velocity depends on the volume change of the whole core rather than of the surface layer alone, we replace $\bar v_{\rm H}$ in Equation \ref{eq:vbar} with the bulk value $\partial V_{\rm core}/\partial M_{\rm H_2}$ obtained from the equation of state of the molten core, where $M_{\rm H_2}$ is the mass of hydrogen in the melt.  This formulation effectively averages the pressure-dependent partial specific volume over the depth of the magma ocean.

This moving phase change horizon (Equation \ref{eq:rfdot}) is analogous to the  moving evaporation front of \citet{Young1998-3109},  but with the Danckwerts-based dissolution flux \eqref{eq:JH} replacing the Hertz--Knudsen sublimation flux, and the boundary advancing outward and consuming mass rather than receding into the evaporating phase with mass loss.

The consolute surface, the radius at which the local temperature reaches the solvus crest (Figure \ref{fig:phase_diagram}),
$\mathcal{G}\equiv T-\Tcrit(P)=0$, is itself a moving thermodynamic level set,
\begin{equation}
\dot r_b=-\frac{\partial_t\mathcal{G}}{\partial_r\mathcal{G}},
\label{eq:rb}
\end{equation}
that moves inward as the planet cools. The engulfment front $r_\varphi$ chases it, and the two meet when the core-envelope interface finally cools through the crest of the binodal and the miscible region closes. The consolute surface is a convenient tracer. The actual concentration of H$_2$ on the melt side of the binodal is is specified by the composition-dependent binodal $T$ (Figure \ref{fig:phase_diagram}), and in practice defines the concentration at the engulfment front once equilibrium is attained there.

\section{Buoyancy Limitations on Ingassing}
\label{sec:buoyancy}
Because dissolving hydrogen into silicate melts lowers their densities, an objection to
large-scale infusion of H$_2$ from a hydrogen-rich envelope into the underlying molten
core in a growing gas-dwarf sub-Neptune is that a gravitationally stable, under-dense
boundary layer will form at the top of the melt and block further ingassing
\citep[e.g.,][]{Modirrousta-Galian_2025}. This picture assumes the boundary-layer melt
is never swept back down. Here it is shown instead that the same surface cooling that keeps the
magma ocean convecting continually renews its surface, so that ingassing is \emph{limited}
by the buoyancy of the low-density layer but never \emph{eliminated} by it: the flux is
confined between zero and a cooling-prescribed maximum $J_{\rm crit}$, and stays finite for as
long as the ocean convects. Paradoxically, surface renewal both drives hydrogen ingassing and limits it, since the dissolved low-density hydrogen-bearing melt weakens the convective overturn that creates it.

Surface heat loss drives the process. A magma ocean beneath an envelope is
cooled from above; heat lost at the surface makes cold, dense parcels that sink, and this
drives the overturn that renews the surface (Equation~\ref{eq:danckwerts}) and carries
hydrogen inward. Nothing else stirs the ocean (effects of rotation are considered below), so when the cooling stops the
overturn, and the ingassing it powers, stops with it. The strength of the drive is the
rate of buoyancy production resulting from the intrinsic (escaping) heat flux $\Fint$ through the
thermal buoyancy flux
\begin{equation}
B_T=\frac{g\,\alpha\,\Fint}{\rho\,c_P},
\label{eq:BT}
\end{equation}
where $\alpha$ is the melt expansivity (K$^{-1}$), $c_P$ its specific heat
(J\,kg$^{-1}$\,K$^{-1}$), and $\Fint$ the escaping heat flux (W\,m$^{-2}$). $B_T$ is the rate at which cooling feeds buoyant energy to the flow per unit mass, and in
steady convection it is balanced by turbulent dissipation. A depth-spanning cell of speed
$\vconv$ and size $d_{\rm ocean}$ overturns in a time
$t_{\rm over}\sim d_{\rm ocean}/\vconv$ and carries kinetic energy $\sim\vconv^2$ per unit
mass, so it dissipates at
\begin{equation}
\epsilon_{\rm turb}\sim\frac{\vconv^2}{t_{\rm over}}\sim\frac{\vconv^3}{d_{\rm ocean}} .
\end{equation}
Equating supply and dissipation, $B_T\sim\vconv^3/d_{\rm ocean}$, gives the
free-convection (mixing-length) velocity
\begin{equation}
\vconv\sim\left(B_T\,d_{\rm ocean}\right)^{1/3}.
\label{eq:vconv}
\end{equation}
Although $B_T$ is a surface boundary-layer quantity, in steady convection this same
buoyancy flux sets the velocity of the overturn, and hence the renewal rate
$s=\vconv/d_{\rm ocean}$. This surface control follows
because here we assume cooling is the sole buoyancy source: absent basal or internal
heating, the interior is nearly adiabatic (marginally stable) and convects only in response to
cooling driven from the surface, so neutralizing the surface buoyancy would arrest the
whole overturn motion. Ingassing at the surface opposes this drive. Because hydrogen diffuses only modestly slower than heat here (Lewis number $\Le=\kappath/\Dhyd\approx5$--$8$), the compositional sublayer is only
somewhat thinner than the thermal layer, $\delta_{\theta}$, with $\dc=\dth\,\Le^{-1/3}\approx0.5\,\dth$. The 
thermal and compositional anomalies therefore occupy essentially the same descending surface material
and their buoyancies combine as a single net buoyancy flux. With the
compositional expansion coefficient $\beta_X=-\rho^{-1}(\partial\rho/\partial X)>0$
(Equation~\ref{eq:vbar}), ingassing at flux $J_{\rm H}$ supplies a stabilizing compositional
buoyancy flux $B_X=g\,\beta_X J_{\rm H}/\rho$, so the overturn is now powered by the net buoyancy flux,
\begin{equation}
B_{\rm net}=B_T-B_X ,
\label{eq:BX}
\end{equation}
and dissolving hydrogen slows the very circulation that delivers it. The two fluxes
balance, $B_{\rm net}=0$, at the critical flux
\begin{equation}
J_{\rm crit}=\frac{B_T\,\rho}{g\,\beta_X}=\frac{\alpha\,\Fint}{\beta_X\,c_P} ,
\label{eq:Jcrit}
\end{equation}
the ingassing flux whose stabilizing buoyancy would exactly cancel the thermal
drive, and, like $B_T$, a quantity fixed by the cooling flux $\Fint$.
Because the overturn is driven by 
the cooling, it exists only while its net budget is positive, $B_{\rm net}>0$, so that
$\vconv\propto B_{\rm net}^{1/3}>0$. The compositional buoyancy can erode the thermal drive
but cannot cancel it as long as convection is extant. Were $J_{\rm H}$ to reach $J_{\rm crit}$, $B_{\rm net}$ and the overturn it drives would vanish, yet that overturn is what renews the surface and delivers the hydrogen; the flux that would cancel the drive depends on the convection its cancellation would destroy. 
A stationary system ($B_{\rm net}=0$, $\vconv=0$) is inconsistent with a magma ocean cooled from the top. In summary, for  $\Fint>0$ the overturn persists, $B_{\rm net}$ stays positive, and
$J_{\rm H}$ remains below $J_{\rm crit}$. 

At the same time, as long as the interior is
undersaturated ($\Delta X=X_{\rm surf}-X_{\rm melt}> 0$) a positive flux of hydrogen persists. 
The ingassing flux of hydrogen is thus confined to the non-zero range
\begin{equation}
0<J_{\rm H}<J_{\rm crit},
\label{eq:bound}
\end{equation}
as long as cooling keeps the surface convecting and renewed.
The overall result is that a low-density boundary layer \emph{limits} ingassing, with $J_{\rm H}$ approaching $J_{\rm crit}$, but 
does not block it.  In this sense, and borrowing from the title of the work by \cite{lee2015a}, ``to cool is to [engulf  hydrogen]".

The value of $J_{\rm H}$ within these bounds follows from the feedback. The overturn
velocity responds to the net drive as
\begin{equation}
\frac{\vconv}{\vconv^{0}}=\left(\frac{B_{\rm net}}{B_T}\right)^{1/3}
=\left(1-\frac{J_{\rm H}}{J_{\rm crit}}\right)^{1/3},
\label{eq:vratio}
\end{equation}
with $\vconv^{0}\propto(B_T d_{\rm ocean})^{1/3}$ being a reference thermal-only value
(Equation~\ref{eq:vconv}). 
Because $s=\vconv/d_{\rm ocean}$ and
$\kex=\sqrt{\Dhyd\,s}\propto\sqrt{\vconv}$, the stabilizing compositional buoyancy reduces the renewal coefficient according to,
\begin{equation}
\frac{\kex}{\kex^{0}}=\left(1-\frac{J_{\rm H}}{J_{\rm crit}}\right)^{1/6},
\label{eq:kthrottle}
\end{equation}
and since $J_{\rm H}=\rho\,\kex\,\Delta X=J_{\rm H}^{0}(\kex/\kex^{0})$ with
$J_{\rm H}^{0}=\rho\,\kex^{0}\Delta X$ being the surface-renewal flux with the overturn driven by the thermal buoyancy only (Equation~\ref{eq:JH}), we obtain the implicit equation for hydrogen flux,
\begin{equation}
J_{\rm H}=J_{\rm H}^{0}\left(1-\frac{J_{\rm H}}{J_{\rm crit}}\right)^{1/6}.
\label{eq:implicit}
\end{equation}
In terms of the buoyancy-flux ratio\footnote{Note the symbol $\mathcal{R}_B$ here is not the Bondi radius.} $\mathcal{R}_B\equiv B_X/B_T = J_{\rm H}/J_{\rm crit}$ and $\Lambda\equiv J_{\rm H}^{0}/J_{\rm crit}$, this self-consistency condition for the hydrogen flux is
$\mathcal{R}_B=\Lambda\,(1-\mathcal{R}_B)^{1/6}$, or equivalently, 
\begin{equation}
\mathcal{R}_B^{\,6}+\Lambda^{6}\mathcal{R}_B-\Lambda^{6}=0,\qquad 0\le \mathcal{R}_B<1 ,
\label{eq:RB}
\end{equation}
from which the single root permits calculation of the self-consistent flux from  $J_{\rm H}=\mathcal{R}_B\,J_{\rm crit}$---the reduced
flux that infuses the interior with hydrogen (Equation~\ref{eq:load}) and advances the front
(Equation~\ref{eq:rfdot}). For any finite drive the root satisfies $\mathcal{R}_B<1$, i.e.\
$J_{\rm H}<J_{\rm crit}$, in keeping with the bound (Equation~\ref{eq:bound}). Two limits illustrate the behavior. For weak drive, $\Lambda\ll1$, $\mathcal{R}_B\approx\Lambda$ and $J_{\rm H}\approx J_{\rm H}^{0}$: ingassing
runs at the full surface-renewal rate and buoyancy is irrelevant. For strong drive,
$\Lambda\gg1$, $\mathcal{R}_B\to1$ and $J_{\rm H}\to J_{\rm crit}$; the flux is pressed up against the convective
ceiling however large the chemical drive. A dry interior beneath a thick, highly soluble H$_2$-rich
envelope sits in this regime, with ingassing at close to the buoyancy imposed ceiling.

Two features of $J_{\rm crit}$ follow from its dependence on $\Fint$. First, as noted previously, ingassing is
\emph{cooling-limited}: since $J_{\rm crit}\propto\Fint$, a perfectly insulated,
non-convecting ocean has $J_{\rm crit}\to0$ and engulfment ceases, so the process is tied
to the molten, actively cooling lifetime rather than to solubility or diffusivity. Second,
the rate is determined by the envelope through $\Fint$: as atmospheric escape thins the overlying
envelope and $\Fint$ rises, $J_{\rm crit}$ lifts and ingassing accelerates. 

The hydrogen flux is stabilized by negative feedbacks. A flux above it raises
the compositional buoyancy $B_X$, lowers the net drive $B_{\rm net}$, and weakens the
overturn, pulling $J_{\rm H}$ back down. A flux below it strengthens the overturn and pushes it
back up. The flux therefore self-regulates to $\mathcal{R}_B\,J_{\rm crit}$, just below the cap,
approaching $J_{\rm crit}$ itself where $\mathcal{R}_B\to1$. The circumstances considered here sit near this limit through most of the engulfment window, because the thermal-buoyancy driven uptake far exceeds the critical limit imposed by the opposing, stabilizing buoyancy, so $\Lambda=J_{\rm H}^{0}/J_{\rm crit}\gg1$.  

Until now we have not considered explicitly the condition $B_{\rm net}\le0$. We do so here to illustrate that a stagnant low-density layer should be a transient condition. We
consider the case where the compositional buoyancy nullifies the thermal
drive, $B_{\rm net}\le0$, overturn stops, and surface renewal ceases.
Hydrogen enters the interface by molecular diffusion only, at a flux
$J_{\rm diff}(t)=\Delta C\sqrt{D/(\pi t)}$. Since the concentrations $X_{\rm surf}$
and $X_{\rm melt}$ are fixed by the overlying envelope and the magma ocean at depth, respectively, addition of hydrogen requires that the stagnant layer grows in depth with time. 
As the low-density region grows in depth, $\delta=\sqrt{\pi Dt}$, the diffusive flux
falls since $J_{\rm diff}=\Delta C\sqrt{D/(\pi t)}=D\,\Delta C/\delta$,
where $\Delta C=\rho(X_{\rm surf}-X_{\rm melt})$. However, the critical flux
$J_{\rm crit}=\alpha F_{\rm int}/(\beta_X c_P)$, prescribed by the cooling flux, changes much less.

To see this, we show that the the thermal and compositional contributions to total buoyancy flux respond differently as a stagnant layer grows. The
stabilizing term falls with increasing thickness, $B_X\propto J_{\rm diff}=D\,\Delta C/\delta$,
whereas $B_T$ is proportional to $F_{\rm int}$ that is generally controlled by the surface temperature and the properties of the envelope. 

The stagnant layer could potentially influence $F_{\rm int}$ by altering the conductive resistance to heat flow in series with the normally dominant resistance of the envelope. The thermal resistance across the evolving stagnant layer is $\delta/k$ where
$k=\rho c_P\kappa$, with thermal diffusivity $\kappa\approx10^{-6}\,\mathrm{m^2\,s^{-1}}$. The envelope's differential thermal resistance per unit area, evaluated about the no-layer state, is
$R_{\rm env} = \left(dF_{\rm int}/dT_{\rm surf}\right)^{-1}
= T_{\rm surf}/(nF_0)$,
where the derivative is with respect to a small change in the magma surface temperature
$T_{\rm surf}$, $F_0$ is the intrinsic flux in the absence of the stagnant layer, and
$n = d\ln F_{\rm int}/d\ln T_{\rm surf}$ is evaluated at that state
($n=4$ for radiation directly to space, but differs in the presence of an atmosphere).
This is a local linear approximation that applies when the temperature change is small
enough that $n$ and the envelope properties remain approximately constant. 

Treating the stagnant layer and envelope as thermal resistances in series, the heat flux is
$F_{\rm int}\approx F_0 R_{\rm env}/(R_{\rm env}+\delta/k)
= F_0/(1+\delta/\delta_T)$. The critical  thickness $\delta_T=kR_{\rm env}$ is defined by setting the stagnant-layer
resistance, $\delta/k$, equal to the envelope resistance, $R_{\rm env}$; at
$\delta=\delta_T$, the two resistances are equal. For $\delta\ll\delta_T$, the envelope limits the heat flux. For
$\delta\gg\delta_T$, the stagnant layer dominates. When $\delta\ll\delta_T$, $B_T$ is essentially constant
while $B_X$ falls as $1/\delta$. When $\delta \gg \delta_T$, $F_{\rm int}$, and with it $J_{\rm crit}$, also falls as
$1/\delta$, so the thermal and compositional buoyancy contributions decline together. Overturn resumes once the diffusive hydrogen flux becomes too small to
offset the cooling, $J_{\rm diff}<J_{\rm crit}$ (equivalently $B_{\rm net}>0$), which
occurs when the layer reaches the critical thickness $\delta_\ast=D\,\Delta C/J_{\rm crit}$,
at $t_\ast=\delta_\ast^2/(\pi D)=D\,\Delta C^{2}/(\pi J_{\rm crit}^{2})$. Overturn therefore resumes provided that 
\begin{equation}
\frac{\delta_\ast}{\delta_T}=\frac{n\,D\,\Delta C\,\beta_X c_P}{\alpha\,k\,T_{\rm surf}}<1 ,
\label{eq:deltas}
\end{equation}
a condition that is independent of cooling rate.

Relevant material properties indicate that the ratio in Equation \ref{eq:deltas} is significantly less than unity, limiting the duration of a stagnant layer.  The envelope $n$ parameterizes how steeply the envelope's
heat flux varies with surface temperature. A hotter surface places the rcb at lower
pressure, where the gas is less dense and less opaque. Combining radiative diffusion at the
rcb, $F_{\rm int}\propto1/(\kappa_{\rm R}P_{\rm rcb})$, with the pressure dependence of the
Rosseland opacity, approximated as $\kappa_{\rm R}\propto P^{\,a}$ with $a\approx0.68$, and
with the adiabat, $P_{\rm rcb}\propto T_{\rm surf}^{-1/\nabla_{\rm ad}}$, results in
$F_{\rm int}\propto T_{\rm surf}^{\,n}$ with $n\approx(1+a)/\nabla_{\rm ad}$. For an ideal
diatomic H$_2$ envelope, $\nabla_{\rm ad}=2/7$ and $n\approx6$. With $n=6$ and representative magma-ocean values,
$D=10^{-8}\,\mathrm{m^2\,s^{-1}}$, $\Delta C=54\,\mathrm{kg\,m^{-3}}$, $\beta_X=5$,
$c_P=1200\,\mathrm{J\,kg^{-1}\,K^{-1}}$, $\alpha=3\times10^{-5}\,\mathrm{K^{-1}}$,
$k\approx1.6\,\mathrm{W\,m^{-1}\,K^{-1}}$ and $T_{\rm surf}=3000$~K, we obtain
$\delta_\ast/\delta_T\approx0.14$. The condition for overturn is therefore met with a margin of
$\sim$7, and the reduction in $F_{\rm int}$ caused by the stagnant layer lengthens $t_\ast$ by only a factor of $(1-\delta_\ast/\delta_T)^{-2}\approx1.3$.

We can estimate the duration of the stagnant condition for representative
hydrogen-bearing magma-ocean properties of $\rho\approx1350\,\mathrm{kg\,m^{-3}}$,
$D\approx10^{-8}\,\mathrm{m^2\,s^{-1}}$, $\Delta X_s\approx0.04$, $\beta_X\approx5$,
and $F_{\rm int}\approx0.2\,\mathrm{W\,m^{-2}}$, yielding
$J_{\rm crit}\approx10^{-9}\,\mathrm{kg\,m^{-2}\,s^{-1}}$ and
$\delta_\ast\approx540~\mathrm{m}$. Therefore,
\begin{equation}
  t_\ast=\frac{D\,\Delta C^{2}}{\pi J_{\rm crit}^{2}}\approx3\times10^{5}~\mathrm{yr},
  \label{eqn:tstar}
\end{equation}
with $\Delta C=\rho\,\Delta X_s\approx54\,\mathrm{kg\,m^{-3}}$.  Since density and hydrogen mass fraction are inversely correlated, $\Delta C=\rho\,\Delta X$
varies by only $\mathcal{O}(1)$ over the plausible range of surface hydrogen concentrations and melt densities, and thus so does $t_\ast$.
This timescale is considerably shorter than the engulfment timescales of $10^7$ to $10^9$ yr (see Section \ref{sec:results}). In treating $X_{\rm melt}$ as fixed here we are being conservative: were the far-field hydrogen concentration to rise as hydrogen accumulates at depth, $\Delta C$, and hence
$J_{\rm diff}$, would fall further, shortening $t_\ast$ and hastening the recovery of overturn.
The stagnant condition is therefore expected to be transient:
diffusion simultaneously builds the boundary and dilutes the gradient driving diffusion until the undiminished cooling
once again dominates, after which the feedback described above returns $J_H$ to
$\mathcal{R}_B J_{\rm crit}$.  We see again, for any finite cooling large enough to matter on a planetary scale, $B_{\rm net}>0$ is inexorable and so is the surface renewal permitting hydrogen engulfment.

A caveat is that this analysis relies on the net buoyancy flux
$B_{\rm net}$, which is exact only for a Lewis number of unity. It therefore cannot
represent double-diffusive behavior by construction. The possibility that the bounding diffusive region comprises instead an oscillatory, double-diffusive layering, rather than the purely transient behavior described here, is examined in \S~\ref{sec:double}. 

\section{Energy Limitations}
\label{sec:energy}
As ingassing loads the surface and $\mathcal{R}_B\to1$, the compositional buoyancy flux grows to nearly
match the thermal buoyancy flux, the net drive $B_{\rm net}=B_T-B_X$ falls toward zero, and the overturn slows to its weakest. This is where double-diffusive layering is most
favored, and where two additional limits not addressed explicitly by the surface-renewal condition
$\mathcal{R}_B<1$ must be checked: whether convection this weak still carries the
mechanical \emph{power} to redistribute hydrogen through the deep ocean, and whether the
composition freezes into a static, layered barrier. For the conditions relevant for sub-Neptune formation, neither blocks ingassing: the convection remains powerful enough, and the layers that form are
transient.

\subsection{Energy of Mixing}
\label{sec:energy_limits}
Incorporating the buoyant surface melt into the denser magma-ocean interior requires mechanical work against the stabilizing compositional buoyancy. This work is supplied primarily by the kinetic-energy flux of the convective flow \citep{Fuentes_2020}. Here, \emph{entrainment} denotes the incorporation of fluid from the stably stratified, low-density surface layer into the underlying convecting region. Subsequent irreversible mixing increases the background gravitational potential energy of the density field. The energy budget question is therefore whether the fraction of the convective kinetic-energy flux available for entrainment and irreversible mixing is sufficient to overcome the stabilizing buoyancy of the compositional stratification.

Redistributing
the ingassing rate $\dot M_{\rm H}=J_{\rm H} A_\varphi$ through a mixed depth $H\!\approx\!d_{\rm ocean}$
requires a power $\dot W_{\rm mix}=C_g\,\beta_X\,g\,H\,|\dot M_{\rm H}|$, where $C_g\simeq\tfrac12$ is
a geometric factor that specifies the average relative depth through which each parcel must be carried. Convection supplies this work from its kinetic-energy throughput, of which only a fraction $\epsilon_{\rm ent}$ goes into entrainment, the remainder being dissipated as heat.  The available power is therefore $\dot W_{\rm avail}=\epsilon_{\rm ent}\,(\alpha g H/c_P)\,L_{\rm conv}$, with $L_{\rm conv}=L_{\rm int}$. Their ratio is independent
of the  mixing depth and gravity,
\begin{equation}
\Pi_E=\frac{\dot W_{\rm mix}}{\dot W_{\rm avail}}
     =\frac{C_g\,\beta_X\,c_P\,|\dot M_{\rm H}|}{\epsilon_{\rm ent}\,\alpha\,L_{\rm conv}},
\label{eq:PiE}
\end{equation}
where $\Pi_E>1$ means entrainment of hydrogen is energy-limited. The largest
energy supportable flux is $J_{\rm H}^{\rm en}=\epsilon_{\rm ent}\,\alpha\,F_{\rm conv}/(C_g\,\beta_X\,
c_P)$ with $F_{\rm conv}=L_{\rm conv}/A_\varphi$, so $\Pi_E=J_{\rm H}/J_{\rm H}^{\rm en}$. Note that dissolution
also releases chemical free energy, corresponding to a chemical power $\dot W_{\rm chem}$ that is at most $10\%$ of $\dot W_{\rm mix}$ and thus a minor contributor to the energy budget. Because the contribution is minor, and because it is released locally at the reaction site, it is neglected here.

Values for $\epsilon_{\rm ent}$ can be estimated from the Prandtl number, $\mathrm{Pr}$, for the melt phase. In
this marginal regime \citet{Fuentes_2020} simulated $\epsilon_{\rm ent}$ and found that
it varies with the Prandtl number such that 
$\epsilon_{\rm ent}$ values of $\approx0.85,\,0.48$, and $0.10$ correspond to
$\mathrm{Pr}$ values of $\approx 0.1,\,1$, and $7$, respectively. The melt Prandtl number $\mathrm{Pr}=\eta\,c_P/k_{\rm th}$ can be estimated from first-principles MgSiO$_3$ viscosities \citep{Karki_Stixrude_2010}. At the hot,
moderate-pressure conditions where surface renewal acts, the magma ocean is a highly mobile phase that is qualitatively liquid water-like, with $\mathrm{Pr} \simeq 1$ to $4$ at 10,000 to 4,000 K, respectively. These properties indicate that $0.5$ is a reasonable intermediate value for $\epsilon_{\rm ent}$ where temperatures are greatly in excess of 4,000 K, as is the case for planets experiencing a ``hot start" (see below). For the sub-Neptunes modeled here, this yields 
$\Pi_E \rightarrow 1$  over the engulfment window. From this analysis, deep mixing, and thus the flux of hydrogen to the magma ocean, operates near the energy limit where $J_{\rm H} \rightarrow J_{\rm H}^{\rm en}$ and $J_{\rm H}^{\rm en} = J_{\rm crit}$ with $\epsilon_{\rm ent} = 0.5$ and $C_g = 1/2$. Including the small chemical free energy of dissolution as an additional available power source lowers the power ratio just below unity ($\Pi_E\lesssim1$), so $J_{\rm H}$ settles
marginally below $J_{\rm crit}$ rather than precisely at it.  

Here it has been assumed that $f_{\rm cool} = L_{\rm conv}/L_{\rm int} = 1$.  Less efficient cooling due to the enveloping disk environment would act like a reduced efficiency factor because of the term $\epsilon_{\rm ent} L_{\rm conv}$, which can be written as $\epsilon_{\rm ent} f_{\rm cool} L_{\rm int}$, in Equation \ref{eq:PiE}. Model calculations show that the energy limit is encountered ($\Pi_E>1$) for $\epsilon_{\rm ent} f_{\rm cool} \lesssim 0.25$, requiring $f_{\rm cool} \approx 0.5$ with our adopted value for $\epsilon_{\rm ent}$ of $0.5$. Such an extreme effect of the disk gas on cooling of the growing planet is unlikely \citep{ginzburg2016a}.

\subsection{Double Diffusive Layering}
\label{sec:double}
The single net drive $B_{\rm net}=B_T-B_X$ assumes the thermal and compositional
buoyancies act on the same parcels, which is exact only where $\Le=1$ and is a reasonable approximation for the actual $\Le$ values of $\approx5$ to $8$. This raises the possibility that the boundary region could organize into a double-diffusive staircase or a stably stratified surface barrier that prevents ingassing.

The expected behavior can be constrained by the simulations of
\citet{Fuentes_2022}. At $\Pr=0.5$, \citet{Fuentes_2022} generally found no discrete secondary
layers. Instead, convective eddies overshot the base of the outer mixed
region and entrained material from the continuously stratified fluid below.
This progressive erosion caused the convective boundary to advance through
the stable composition gradient until the entire domain was mixed. At $\Pr=7$, distinct secondary convective layers formed within the
stably stratified fluid below the downward-advancing base of the outer
convective region. As that outer region deepened, it reached and engulfed
the secondary layers, which therefore did not prevent eventual complete
mixing; the layers were successively engulfed by the advancing convection zone and
the full region ultimately became well mixed. Our estimated
$\Pr\simeq1$--$4$ lies between these two regimes. Moreover,
\citet{Fuentes_2022} adopted $\Le\approx14$, larger than our estimated
$\Le\approx5$ to $8$; the relatively faster compositional diffusion in the
magma ocean should make the maintenance of sharp compositional interfaces
less favorable. Although the existing simulations do not exclude temporary
layer formation at the upper end of the relevant $\Pr$ range, they disfavor a
permanent, impermeable upper boundary layer. Nonetheless, one could interpret our solutions
with $\mathcal{R}_B$ close to unity as potentially containing unresolved transient layering.

\section{Methods}
\label{sec:model}
In order to explore the efficacy of convective surface renewal for hydrogen engulfment, a relatively simple planet model was built, consisting of a molten core, an overlying hydrogen-rich envelope, an enveloping protoplanetary disk with a specified lifespan, followed by evolution after disk dissipation.  The initial condition is taken to be a hydrogen-free core and an as-yet incomplete hydrogen envelope that grows by accretion in the disk.

In each model calculation, the time evolution of a planet of fixed total mass $M_p$ (e.g., $6\,M_\oplus$) and a hydrogen-rich envelope accreted over gaseous disk lifetime $t_{\rm disk}$ ($3$~Myr for the models shown here) is calculated. The state vector is
\begin{equation}
\boldsymbol{\mathcal{S}}(t)=\big(\,\Tpot,\ X_{\rm melt},\ M_{\rm H_2}^{\rm env},\ r_\varphi\,\big),
\label{eq:state}
\end{equation}
where $\Tpot$ is the interior potential temperature that defines the melt adiabat, $X_{\rm melt}$ is the dissolved interior H$_2$ mass fraction, $M_{\rm H_2}^{\rm env}$ is the H$_2$ mass of the envelope, and $r_\varphi$ is the melt/atmosphere hydrogen engulfment front  radius. The envelope helium mass $M_{\rm He}^{\rm env}$ is monitored alongside $\boldsymbol{\mathcal{S}}$, where $M_{\rm atm}=M_{\rm H_2}^{\rm env}+M_{\rm He}^{\rm env}$ denotes the total envelope (atmospheric) mass. Helium is inert with an initial mole fraction $n_{\rm He}/(n_{\rm H_2}+n_{\rm He})=0.15$ and is not soluble in the core in these calculations.  

In \emph{Stage~I}
($t<t_{\rm disk}$) the disk supplies the envelope, building it up from an initial fraction ($0.1$) of the target toward that target envelope mass by filling the envelope from the disk (treated as an infinite supply of H$_2$ and He), including replacement of whatever the interior ingasses from the growing envelope, all while escape is off. In \emph{Stage~II} ($t\ge t_{\rm disk}$) the system is closed, and the atmosphere both drains into the interior and escapes.
 
\subsection{Molten Core}
\label{sec:model-core}
 
The interior is a single well-mixed, convecting reservoir.  The assumption of a homogeneous interior concentration of hydrogen is validated by the convective/diffusive  P\'eclet number of
$\gtrsim10^{12}$.  The core begins hot and hydrogen free in this ``hot start" modeling, with $X_{\rm melt}(0)=10^{-5}$ and
a core surface temperature prescribed by the initial potential temperature, meant to represent a molten core insulated by an atmosphere of modest mass (see below). Our results are not especially sensitive to this choice as this temperature is quickly overwhelmed by accretional heating.  Recalling Equation \ref{eq:load}, the interior acquires hydrogen at the rate,
\begin{equation}
\frac{\dd X_{\rm melt}}{\dd t}=(1-X_{\rm melt})^{2}\,\frac{J_{\rm H}\,A_\varphi}{\Msil},
\label{eq:dXdt}
\end{equation}
with base area $A_\varphi=4\pi R_{\rm core}^2$ evaluated at the moving melt/atmosphere
front, $R_{\rm core}=\max(r_\varphi,R_{{\rm core},0})$ (\S\ref{sec:model-front}), where 
$R_{{\rm core},0}$ is fixed once at the start using the equation of state for the molten interior described by  \cite{Young2025_Differentiation}. In this work, we consider the core to be entirely miscible, comprising a mixture of silicate, iron, and hydrogen.  We leave the modeling of formation and resorption of a metal core as the hydrogen concentration changes \citep[e.g.,][]{Young2025_Differentiation} for future work. This same equation of state is used to calculate the bulk partial specific volume $\bar v_{\rm H}=\partial V_{\rm core}/\partial M_{\rm H_2}^{\rm core}$, the change in core volume per unit mass of dissolved H$_2$, used to calculate the change in radius with time (\S \ref{sec:model-front}). The model is based on a Vinet equation of state for an MgSiO$_3$--H$_2$ melt integrated along the melt adiabat \citep{Young2025_Differentiation} and tabulated as a function of the dissolved hydrogen content and potential temperature.

\subsection{Hydrogen Envelope}
\label{sec:model-atm}

Atmospheric escape is suppressed while the disk is present and permitted thereafter,
such that the H$_2$ mass of the envelope evolves as
\begin{equation}
\frac{\dd M_{\rm H_2}^{\rm env}}{\dd t}=
\begin{cases}
\dot M_{\rm acc}, & t<t_{\rm disk},\\[3pt]
-J_{\rm H}A_\varphi-\dot M_{\rm esc}, & t\ge t_{\rm disk},
\end{cases}
\label{eq:dMatm}
\end{equation}
where the disk source $\dot M_{\rm acc}$ tops the atmosphere up toward its target by exponential relaxation,
\begin{equation}
\dot M_{\rm acc}=\frac{M_{\rm H_2}^{\rm env,tar}-M_{\rm H_2}^{\rm env}}{\tau_{\rm build}},
\label{eq:mdotacc}
\end{equation}
with $\tau_{\rm build}=0.3\,t_{\rm disk}$, chosen so that the envelope reaches its target ($\simeq97\%$) by the time the disk disperses while building up over most of the disk lifetime, as expected for envelope growth limited by cooling and contraction. The target is specified for the total envelope (H$_2$ + He), where $M_{\rm H_2}^{\rm env,tar}$ is the H$_2$ mass. Helium accompanies the hydrogen at a fixed ratio.  In these calculations the envelope starts at $0.1$ of the target values and grows at constant composition. After the disk disperses, helium is lost only by hydrodynamic drag (\S\ref{sec:model-esc}).

The accreting gas deposits heat into the interior according to the energy balance between accretion heating and radiative cooling,
\begin{equation}
\frac{\dd T_{\rm pot}}{\dd t}
   =\frac{\eta_{\rm eff}\,L_{\rm acc}-L_{\rm int}-L_{\rm boil}}{\epsilon_{\rm \langle T\rangle}M_{\rm sil}\,c_P},
\label{eq:Tpot}
\end{equation}
where $T_{\rm pot}$ is the potential temperature for the convecting silicate interior, $c_P=1200\,\mathrm{J\,kg^{-1}K^{-1}}$ is the specific heat capacity of the interior, $M_{\rm sil}$ is the mass of silicate, $L_{\rm int}$ is the intrinsic luminosity radiated through the envelope
(\S~\ref{sec:model-atm}), $L_{\rm boil}$ is the power drawn from stored heat by boil-off escape shortly after disk dispersal (zero otherwise; \S~\ref{sec:model-esc}), and $\epsilon_{\rm \langle T\rangle}$, of order unity, corrects the heat capacity $M_{\rm sil}c_P$ for $\langle T\rangle/T_{\rm pot}$ where mass-weighted temperature $\langle T\rangle$ is $\ge\!T_{\rm pot}$ along the adiabat \cite[e.g.,][]{Barth_2021}. For simplicity, we set $\epsilon_{\rm \langle T\rangle}$ to 1, making our time evolution of the core conservative. The accretion luminosity is the rate at which the incoming hydrogen liberates gravitational energy as it settles onto the interior,
\begin{equation}
L_{\rm acc}=\frac{G M_p}{R_{\rm core}}\left(\dot M_{\rm acc}+\dot M_{\rm ingas}\right),
\label{eq:Lacc}
\end{equation}
combining the atmosphere supply $\dot M_{\rm acc}$ (Eq.~\ref{eq:mdotacc}) with the
mass dissolved into the melt, $\dot M_{\rm ingas}=J_{\rm H} A_\varphi$ (counted only while
ingassing, $J_{\rm H}>0$), released at the interior radius $R_{\rm core}$.
Once the disk disperses the source vanishes and ingassing then draws directly on the
atmosphere (the $-J_{\rm H}A_\varphi$ term above).

The dimensionless efficiency $\eta_{\rm eff}$ is the fraction of the liberated accretion energy retained as interior heat rather than radiated away at the accretion shock, and it is the model's hot-start/cold-start control \citep{Marley2007}. We treat $\eta_{\rm eff}$ as a constant, time-averaged retained fraction.  We take $\eta_{\rm eff}$ to be unity for the models here and discuss its effects below.  Once the disk disperses, accretion ceases and $L_{\rm acc}=0$, so Equation~\eqref{eq:Tpot} reduces to pure radiative cooling, $\dd T_{\rm pot}/\dd t=-L_{\rm int}/(M_{\rm sil}c_P)$.

At each step a quasi-static, self-gravitating, hydrogen-rich gas with a non-ideal equation of state \citep{Chabrier2019} comprises the envelope with the outer boundary being the Bondi radius or the Hill radius, whichever is smaller, 
$R_{\rm out}=\min(R_{\rm Bondi},R_{\rm Hill})$.  
The structure is simplified for this purpose, and consists of two layers, a convective adiabatic lower layer beneath an isothermal radiative upper layer. The convective structure follows from
hydrostatic equilibrium, 
\begin{equation}
\frac{\dd P}{\dd r}=-\rho_{\rm gas}\,g,
\label{eq:hydro}
\end{equation}
mass continuity, 
\begin{equation}
\frac{\dd M}{\dd r}=4\pi r^2\rho_{\rm gas},\
\label{eq:cont}
\end{equation}
and the adiabat integrated outward from the base
\begin{equation}
\frac{\dd\ln T}{\dd\ln P}=\nabla_{\!\rm ad}(T,P),
\label{eq:adiabat}
\end{equation}
with $\rho_{\rm gas}(P,T)$ and $\nabla_{\!\rm ad}(T,P)$ from the \citet{Chabrier2019} equation of state.
The radiative--convective boundary (rcb) is the radius $R_{\rm rcb}\le R_{\rm out}$ at
which the outward-cooling adiabat reaches the rcb temperature $T_{\rm rcb}$.  If it has not
cooled to $T_{\rm rcb}$ by $R_{\rm out}$, the rcb is capped at $R_{\rm rcb}=R_{\rm out}$
(the inflated, disk-supported regime).

The rcb temperature is specified by the analytical gray Schwarzschild
criterion $\nabla_{\!\rm rad}=\nabla_{\!\rm ad}$ in which the Rosseland opacity is
approximated by a local  power law $\kappa\propto P^{a}$ with $a\simeq\tfrac12$, a
representative slope of the opacity near the rcb \citep{freedman2014a}. This power law
as used here is only a weak geometric factor setting the order-unity ratio $T_{\rm rcb}/T_{\rm eq}$. The intrinsic flux and the $\tau=1$ transit radius are evaluated with the full tabulated \citet{freedman2014a} opacities. We have, accordingly

\begin{equation}
T_{\rm rcb}=T_{\rm eq}\left(1-\frac{4\,\nabla_{\!\rm ad}}{1+a}\right)^{-1/4},
\label{eq:Trcb}
\end{equation}
which for a dry-H$_2$ adiabat ($\nabla_{\!\rm ad}=2/7$) and $a\simeq\tfrac12$ gives
$T_{\rm rcb}\simeq1.4\,T_{\rm eq}$. Since
$T_{\rm rcb}$ is only weakly sensitive to $a$ and the rcb lies near $T_{\rm eq}$,
this approximation does not materially affect the results.

Above the rcb, a thin, isothermal radiative layer with $T = T_{\rm eq}$ and extending
to $R_{\rm out}$ is present once the disk no longer supports the envelope. The mass of the radiative layer is estimated from the hydrostatic column above the rcb, $4\pi R_{\rm rcb}^{2}P_{\rm rcb}/g_{\rm rcb}$. It amounts to $\sim\!1$--$6\%$ of $M_{\rm atm}$, increasing as the rcb deepens with cooling, and it determines the transit radius. The base temperature is adiabatic,
$T_{\rm surf}=T_{\rm ad}(P_{\rm surf},\Tpot,X_{\rm melt})$, and $P_{\rm surf}$ is fixed by requiring the mass of the convective layer plus that of the radiative layer to equal the total envelope mass $M_{\rm atm}$ of hydrogen and helium. 

The intrinsic
luminosity in Equation~\eqref{eq:Tpot} is obtained from the intrinsic flux at the rcb,
\begin{equation}
L_{\rm int}=4\pi R_{\rm rcb}^2\,F_{\rm int},
\label{eq:Lint_L}
\end{equation}
where,
\begin{equation}
F_{\rm int}=\frac{16}{3}\,\frac{\sigma\,\nabla_{\!\rm ad}\,g\,T_{\rm rcb}^4}{\kappa_{\rm rcb}P_{\rm rcb}},
\label{eq:Lint}
\end{equation}
with $\kappa$ the Rosseland mean opacity \citep{freedman2014a}.

\subsection{Atmospheric Escape}
\label{sec:model-esc}

Escape is suppressed while the disk supports the envelope, accounting for continuous
atmospheric rebuilding, and is enabled once the disk dissipates at \(t_{\rm disk}\).
Two escape mechanisms are considered: XUV-driven photoevaporation, powered by the star,
and thermally driven hydrodynamic escape, powered by the planet's own heat. The
H$_2$ escape rate is the larger of the two,
\begin{equation}
\dot M_{\rm esc}=\max\!\big(\dot M_{\rm PE},\,\dot M_{\rm th}\big).
\label{eq:mdotesc}
\end{equation}

The energy-limited photoevaporation rate is
\begin{equation}
\dot M_{\rm PE}=\frac{\epsilon\,\pi\,F_{\rm XUV}\,R_{\rm xuv}^3}{G M_p\,K_{\rm tide}}\,\mathcal T,
\label{eq:mdotPE}
\end{equation}
with efficiency $\epsilon=0.16$, tidal correction $K_{\rm tide}=0.8$, absorption radius
$R_{\rm xuv}=f_{\rm xuv}R_{\rm rcb}$ ($f_{\rm xuv}=1$), and stellar flux
\begin{equation}
F_{\rm XUV}=\frac{L_{\rm XUV}}{L_{\rm bol}}\,\frac{L_{\rm bol}}{4\pi a^2},
\end{equation}
where $L_{\rm XUV}/L_{\rm bol}=f_{\rm sat}$ for $t\le t_{\rm sat}$ and
$\propto t^{-\beta}$ afterward ($f_{\rm sat}=10^{-3}$, $t_{\rm sat}=100$~Myr, $\beta=1.2$;
$L_{\rm bol}$ a broken main-sequence law) \citep[values from][]{Young_2026}. A numerical
depletion taper $\mathcal T=M_{\rm H_2}^{\rm env}/(M_{\rm H_2}^{\rm env}+M_{\rm floor})$
is applied to all escape rates, so $\dot M_{\rm esc}\to0$ as the
H$_2$ of the envelope becomes exhausted. $M_{\rm floor}=10^{-6}\,M_{\rm H_2}^{\rm target}$ ($\approx 6\times10^{-7}\,M_\oplus$ for the fiducial planet).

Thermally driven escape is a transonic isothermal Parker wind launched from the
radiative--convective boundary,
\begin{equation}
\dot M_{\rm w}=\pi\,\frac{(GM_p)^2\,\rho_{\rm rcb}}{c_s^{3}}\,
              \exp\!\Big(\tfrac32-\lambda_{\rm rcb}\Big),
\end{equation}
where the isothermal sound speed is set by the equilibrium temperature,
\begin{equation}
c_s^2=\frac{R_{\rm gas}T_{\rm eq}}{\mu},
\end{equation}
with $\mu$ the mean molecular weight (molar mass) of the H$_2$-rich envelope. The gas
density at the rcb follows from the ideal-gas law at the local rcb temperature,
\begin{equation}
\rho_{\rm rcb}=\frac{\mu\,P_{\rm rcb}}{R_{\rm gas}\,T_{\rm rcb}},
\label{eq:rhorcb}
\end{equation}
and the escape parameter at the rcb, the ratio of gravitational binding to thermal
energy, is
\begin{equation}
\lambda_{\rm rcb}=\frac{GM_p}{c_s^{2}R_{\rm rcb}}.
\end{equation}
The wind becomes sonic at $r_{s}=GM_p/2c_s^{2}$, and the factor
$\exp(\tfrac32-\lambda_{\rm rcb})$ is the isothermal density ratio between the rcb and $r_s$.

The wind operates in two regimes, distinguished by the energy source for lifting the gas out of
the gravitational potential well. In the sustained, luminosity-limited regime (core-powered mass loss),
the outflow cannot exceed what the intrinsic luminosity can lift \citep{gupta2019a},
\begin{equation}
\dot M_{\rm E}=\frac{L_{\rm int}\,R_{\rm rcb}}{GM_p}.
\end{equation}
Immediately after disk dispersal, however, a hot envelope with small
$\lambda_{\rm rcb}$ can instead draw on its stored heat and lose mass at a 
rate corresponding to sonic outflow through the rcb, $\dot M_{\rm s}=4\pi R_{\rm rcb}^2\rho_{\rm rcb}c_s$
\citep[spontaneous boil-off;][]{owen2016a, ginzburg2016a}. Boil-off occurs when the wind would remove the envelope faster than the envelope can
cool and contract \citep{owen2016a, ginzburg2016a}. At dispersal the envelope's cooling
time is comparable to its age, $\sim t_{\rm disk}$, so the criterion is
$\tau_{\rm loss}<t_{\rm disk}$, with $\tau_{\rm loss}=M_{\rm H_2}^{\rm env}/\dot M_{\rm w}$.
This criterion is applied as a smooth step,
\begin{equation}
w=\big[1+(\tau_{\rm loss}/t_{\rm disk})^4\big]^{-1},
\end{equation}
where the exponent $n=4$ is a
numerical smoothing parameter that avoids a
discontinuity in the evolution equations. Because
$\tau_{\rm loss}\propto e^{\lambda_{\rm rcb}}$, the step is sharp in $\lambda_{\rm rcb}$:
$w$ falls from $\approx1$ to $\approx0$ over less than one unit of $\lambda_{\rm rcb}$ near
$\lambda_{\rm rcb}\approx25$, and the exponent and the precise choice of timescale have
little effect on the results. The two regimes give
\begin{equation}
\dot M_{\rm CPML}=\min\!\big(\dot M_{\rm w},\,\dot M_{\rm E}\big)\,\mathcal T
\label{eq:cpml}
\end{equation}
for core-powered mass loss and
\begin{equation}
\dot M_{\rm boil}=w\,\min\!\big(\dot M_{\rm w},\,\dot M_{\rm s}\big)\,\mathcal T
\label{eq:boiloff}
\end{equation}
for boil-off, and the thermally driven rate is the larger of the two,
\begin{equation}
\dot M_{\rm th}=\max\!\big(\dot M_{\rm CPML},\,\dot M_{\rm boil}\big).
\label{eq:mdotth}
\end{equation}
Escape in excess of $\max(\dot M_{\rm PE},\dot M_{\rm CPML})$ requires the specific energy
$GM_p/R_{\rm rcb}$, and the corresponding power, $L_{\rm boil}$, is subtracted from the interior energy balance (Equation~\ref{eq:Tpot}). The resulting cooling contracts
the envelope, raises $\lambda_{\rm rcb}$ and returns the wind to the luminosity-limited
regime within $\sim$0.1~Myr.

Hydrogen is conserved such that
$\dd[M_{\rm H_2}^{\rm env}+M_{\rm H_2}^{\rm core}]/\dd t=\dot M_{\rm disk}-\dot M_{\rm esc}$,
where $M_{\rm H_2}^{\rm core}=X_{\rm melt}M_{\rm sil}/(1-X_{\rm melt})$ is the H$_2$
dissolved in the core, with disk source $\dot M_{\rm disk}=\dot M_{\rm acc}+J_{\rm H}A_\varphi$
in Stage~I and zero thereafter.

\subsection{Helium}
Helium in the envelope is treated as a chemically inert, relatively transparent diluent with 
a solar concentration relative to hydrogen during accretion from the disk gas. Although unreactive, He affects chemistry indirectly by contributing its weight to the hydrostatic
pressure that affects the melt equation of state and the consolute temperature of the binodal phase boundary. The Rosseland mean opacity of the envelope is a mass-weighted mix of the solar-composition and pure-helium values, $\kappa=w_{\rm H_2}\,\kappa_{\odot}+(1-w_{\rm H_2})\,\kappa_{\mathrm{He}}$,
where here the weight $w_{\rm H_2}$ is the H$_2$ mole fraction normalized to its solar
value and with an upper bound of unity; the solar $\kappa_\odot$ is recovered for a solar composition, and the opacity shifts toward the helium value as hydrogen is depleted.
The helium end-member is a gray Rayleigh-scattering estimate, about eight orders of magnitude
smaller than solar and about $15$ times weaker than the H$_2$ Rayleigh opacity
\citep{DalgarnoWilliams1962,SneepUbachs2005}.

Helium raises the mean molecular weight of the envelope, lowering the
isothermal-wind sound speed and exponentially
suppressing hydrogen loss. Helium itself is removed by
hydrodynamic drag once the hydrogen wind is strong enough. The effect is invoked
using the two-component crossover-mass formulation of \cite{HuntenPepinWalker1987}, which determines the helium loss rate $\dot M_{\rm He}$ that depletes $M_{\rm He}^{\rm env}$ in Stage~II.
Strong hydrodynamic escape drags He nearly unfractionated from H$_2$ whereas slow escape
progressively enriches the envelope in He in these calculations.

\subsection{Engulfment Front}
\label{sec:model-front}
 
The flux of Hydrogen crossing the magma ocean-envelope interface is calculated as follows. The surface concentration driving ingassing depends on whether, for a given $P$, the surface temperature lies above the binodal or on the binodal, such that
\begin{equation}
X_{\rm surf}=
\begin{cases}
X_{\rm bulk}, & T_{\rm surf}\ge T_{\rm crit}(P_{\rm H_2}),\\[3pt]
x_{\rm sat}(T_{\rm surf},P_{\rm H_2}), & T_{\rm surf}<T_{\rm crit}(P_{\rm H_2}),
\end{cases}
\label{eq:Xsurf}
\end{equation}
i.e., bulk H$_2$ concentration for supercritical/miscible vs.\ the melt-side binodal composition. Below the consolute temperature, if the bulk hydrogen content lies below the melt-side binodal, the melt--envelope system is single phase at the interface and $X_{\rm surf}=X_{\rm bulk}$. The transition from $X_{\rm bulk}$ to $x_{\rm sat}$ at the consolute temperature is spread over a narrow interval, $0.998<T_{\rm surf}/T_{\rm crit}<1$, for numerical stability. The purely thermally-driven Danckwerts flux is first calculated from
\begin{equation}
\begin{aligned}
J_{\rm H}^{0}&=\rho\,k_{\rm L}\,(X_{\rm surf}-X_{\rm melt}),\\
k_{\rm L}&=\sqrt{D_{\rm H_2}\,s},\\
s&=v_{\rm conv}/R_{\rm core},
\end{aligned}
\label{eq:JHplus}
\end{equation}
with chemical diffusivity $D_{\rm H_2}=D_{\rm self}\Phi$ ($D_{\rm self}$ an Arrhenius
self-diffusivity, $\Phi$ the Darken factor from the binodal's Gibbs free energy model) and the renewal length taken as the magma ocean depth. The effects of rotation are accounted for in the overturn velocity, such that
\begin{equation}
\begin{aligned}
v_{\rm conv}&=\min(u_{\rm free},u_{\rm rot}),\\
u_{\rm free}&=(B_{T}\,R_{\rm core})^{1/3},\\
\end{aligned}
\label{eq:vall}
\end{equation}
and,
\begin{equation}
u_{\rm rot}=(B_{T}/\Omega)^{1/2},
\end{equation}
where both $u_{\rm free}$ and $u_{rot}$ are vertical components of convective velocity; it is the vertical component that replenishes the surface.  Rotation periods are assumed to reflect tidal locking for orbital periods shorter than 10 days and Neptune's rotation period for orbital periods longer than 10 days.  Fast rotation ($u_{\rm rot}<u_{\rm free}$)
suppresses $v_{\rm conv}$ and hence $k_{\rm L}$. 
Where ingassing obtains, $J_{\rm H}^{0}>0$, the hydrogen flux is reduced by the compositional-buoyancy using $\mathcal{R}_B$ of \S~\ref{sec:buoyancy}: the model computes
$\Lambda=J_{\rm H}^{0}/J_{\rm crit}$ with $J_{\rm crit}=\alpha F_{\rm int}/(\beta_X c_P)$, solves
$\mathcal{R}_B=\Lambda(1-\mathcal{R}_B)^{1/6}$ numerically, and determines the flux using
\begin{equation}
J_{\rm H}=
\begin{cases}
\min\!\left(\mathcal{R}_B\,J_{\rm crit},\ J^{\rm en}_{H}\right), & J_{\rm H}^{0}>0,\\[3pt]
J_{\rm H}^{0}, & J_{\rm H}^{0}<0.
\end{cases}
\label{eq:JH_cases}
\end{equation}
The ingassing flux ($J_{\rm H}^{0}>0$) is the smaller of two ceilings: the
buoyancy-limited value $\mathcal{R}_B\,J_{\rm crit}$ (\S\ref{sec:buoyancy}) and the
energy-limited value $J^{\rm en}_{\rm H}$ (\S\ref{sec:energy_limits}); for the fiducial
$\epsilon_{\rm ent}$ and $C_g$ these two ceilings coincide numerically. Neither ceiling
applies to degassing ($J_{\rm H}^{0}<0$), which proceeds at the full surface-renewal rate
$J_{\rm H}^{0}$. This is because dissolving hydrogen must be carried downward against the stabilizing, H$_2$-enriched surface layer it creates, the work supplied by convection, whereas
degassing of hydrogen leaves behind a denser, hydrogen-depleted residue that overturns spontaneously, releasing gravitational potential energy rather than consuming it.

The resulting flux drives the interior evolution at each time step. The dissolved inventory of hydrogen grows according to Equation \ref{eq:dXdt}, and advancement of the front can be written as 
\begin{equation}
\dd r_\varphi/\dd t=\bar v_{\rm H}\,\dot M_{\rm H}/A_\varphi, 
\label{eq:drphi}
\end{equation}
where $\dot M_{\rm H}=J_{\rm H}A_\varphi$ and the partial specific volume $\bar v_{\rm H}=\partial V_{\rm core}/\partial M_{\rm H_2}^{\rm core}$ is obtained from the molten core equation of state.

The interior gravity, area $A_\varphi=4\pi R_{\rm core}^{2}$, and heat flux are re-evaluated
self-consistently at each time step. The position of the consolute surface $r_b$ (where $T=T_{\rm crit}(P_{\rm H_2})$) is computed from the envelope structure at each time step for reference.

\subsection{Hydrogen Diffusivity}
\label{sec:diffusivity}

The degassing rate is based on hydrogen diffusivity $\Dhyd$ (for consistency of notation we refer to the diffusing species as H$_2$ but recognize that the actual diffusivities are atomic in the melt phase). \cite{Karki_2010} report first-principles
\emph{hydrogen} diffusion in hydrous MgSiO$_3$ over $2000$--$6000\,\mathrm K$ and
$0$--$150\,\mathrm{GPa}$. The diffusivity follows an Arrhenius law with a small
pressure (activation-volume) term,
\begin{equation}
  \Dhyd(P,T) = D_0\,
  \exp\!\left(-\,\frac{E_a + P\,V_a}{R\,T}\right),
  \label{eq:karki}
\end{equation}
with (low-pressure fit) $D_0=341\times10^{-9}\,\mathrm{m^2\,s^{-1}}$,
$E_a=60\,\mathrm{kJ\,mol^{-1}}$, and a per-particle activation volume
$V_a=0.27$\AA$^3$ (converted to molar). The Darken thermodynamic factor
$\Phi$ is used to convert the diffusion coefficient to reflect the chemical potential driving force for diffusion. The pressure term is negligible at
the $\sim$few-GPa surface conditions ($P V_a$ is $<1\%$ of $E_a$), so here $\Dhyd$ is essentially dependent only on temperature.

\subsection{Fiducial Planet and Initial Conditions}
For our fiducial case we adopt an orbital period of 50 days around a solar-like star. Other choices would be reasonable, but this one has several advantages. The intrinsic occurrence of sub-Neptunes rises steeply with orbital period out to $P\sim10$ days and is roughly flat per logarithmic interval in period beyond that, out to the limit of current survey completeness at $\sim$100--300 days \citep[e.g.,][]{howard2012a,Fressin2013,petigura2018a}. A period of $\sim$50 days lies near the logarithmic midpoint of this range and so is broadly representative of the population. In addition, at these separations sub-Neptunes outnumber stripped super-Earths, whereas within $\sim$10--20 days the population is shaped substantially by atmospheric loss \citep[e.g.,][]{owen2017a, gupta2019a}. A 50-day orbit therefore allows the effects of engulfment and degassing to be examined with less influence from concurrent envelope stripping.

One can place the initial conditions for the fiducial example reported here in context by utilizing existing closures to estimate the initial potential temperature for the core and the target mass fraction of atmosphere at the end of the protoplanetary disk gas phase. The initial potential temperature of the interior is specified by the gravitational binding energy for the core mass $M_{\rm core}$ and radius $R_{\rm core}$ given 
$E_{\rm grav}=(3/5)\,GM_{\rm core}^{2}/R_{\rm core}$. A fraction $\eta_{\rm core}$ of this
energy is retained as heat (the remainder being radiated away during accretion), producing a mean
temperature rise $\Delta T = (3/5)\,\eta_{\rm core}\,GM_{\rm core}/(R_{\rm core}\,c_P)$, so
that the mass-averaged interior temperature is $T_{\rm bulk}=T_0+\Delta T$, with $c_P$ the
silicate-melt heat capacity and $T_0$ a small reference temperature. The potential temperature $T_{\rm pot}$ at zero pressure can be obtained from the mass-averaged temperature 
along an isentrope using
$T_{\rm pot}=T_{\rm bulk}\,(\rho_0/\rho_{\rm int})^{\gamma}$, where for this estimate the Gr\"uneisen parameter $\gamma$ for
the melt is assumed to be constant, $\rho_0$ is the low-pressure reference density, and
$\rho_{\rm int}\simeq M_{\rm core}/((4/3)\pi R_{\rm core}^{3})$ the average compressed
interior density. We choose a value for $\eta_{\rm core}$ of $0.075$, consistent with previous estimates for accretion of molten rocky cores \citep{Boujibar_2020}. The target atmospheric mass fraction $M_{\rm atm}^{\rm tar}/M_p$, which includes both H$_2$ and He, at the end of the disk gas phase is obtained from Equation~(18) of \cite{ginzburg2016a}. None of these choices impacts the fundamental results shown here.

\subsection{Time Stepping}
At each time step the equations comprising the model are evaluated in five stages, as follows. (1)~The state vector $\boldsymbol{\mathcal{S}}(t)$ (Equation~\ref{eq:state}) specifies the interior potential temperature, the dissolved hydrogen mass fraction, the H$_2$ mass of the envelope, and the radius of the core surface, which is also the position of the engulfment front. The He mass of the envelope is tracked in parallel. (2)~From these, the envelope structure is solved. The pressure at the base of the envelope is adjusted iteratively, with the surface temperature following the melt adiabat at each trial pressure, until the envelope, the convective layer integrated outward to the rcb (Equations~\ref{eq:hydro}--\ref{eq:adiabat} and \ref{eq:Trcb}) plus the radiative layer above it, contains the total envelope mass $M_{\rm atm}$ of H$_2$ and He. This gives the surface temperature and pressure and the intrinsic luminosity (Equation~\ref{eq:Lint_L}). (3)~The surface conditions then determine the hydrogen flux across the melt--envelope interface, through the surface composition (Equation~\ref{eq:Xsurf}), the diffusivity (Equation~\ref{eq:karki}), and the convective velocity (Equation~\ref{eq:vall}), which give the surface-renewal flux (Equation~\ref{eq:JHplus}), limited during ingassing by the buoyancy and energy ceilings (Equation~\ref{eq:JH_cases}). (4)~Next, the source and sink rates at which the envelope gains and loses mass are evaluated. While the disk is present, the disk supply (Equation~\ref{eq:mdotacc}) replaces what is ingassed into the interior, escape is suppressed, and the settling gas releases the accretion luminosity that heats the interior (Equation~\ref{eq:Lacc}). After the disk disperses, hydrogen crosses the interface in both directions, and the net flux determines whether the envelope loses H$_2$ to the interior (ingassing) or gains it from the interior (degassing); the envelope also loses H$_2$ and entrained He to escape by photoevaporation and thermally driven outflow, including transient boil-off (Equations~\ref{eq:mdotesc}--\ref{eq:mdotth}). (5)~Finally, the interface flux, the source and sink rates, the intrinsic and accretion luminosities, and the energy drawn by boil-off together give the time derivatives of the interior temperature (Equation~\ref{eq:Tpot}), the dissolved hydrogen (Equation~\ref{eq:dXdt}), the envelope H$_2$ mass (Equation~\ref{eq:dMatm}), and the engulfment front (Equation~\ref{eq:drphi}), allowing $\boldsymbol{\mathcal{S}}(t)$ to be advanced. The consolute radius $r_b$ and the transit radius are computed from the envelope structure as outputs.

The state is integrated over time with adaptive solvers for stiff systems from the \texttt{scipy.integrate} module of the Python SciPy library \citep{Virtanen2020}, using the LSODA method and switching to the implicit BDF method once the surface temperature approaches the consolute temperature where the exchange of hydrogen between melt and envelope becomes stiff. The integrator tracks the cumulative masses so the open-system mass (\S\ref{sec:model-esc}) balances to integrator tolerance.

\vspace{1em}
\section{Results}
\label{sec:results}
\begin{figure*}[!t]
\centering
\includegraphics[width=0.75\textwidth]{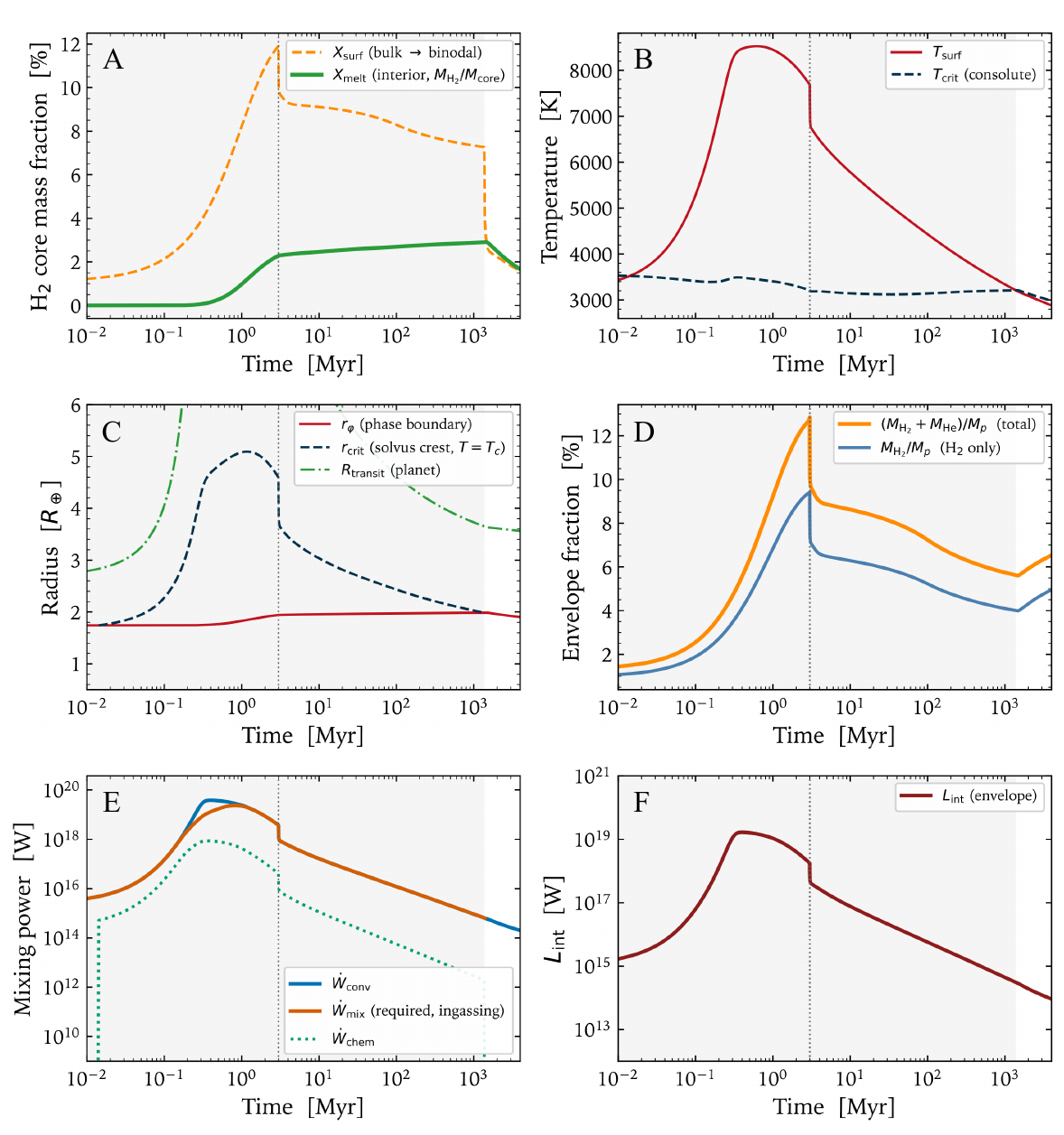}
\caption{Time evolution of a 6$M_{_\oplus}$ planet with a 50 day orbital period around a solar-like star that accumulates a 13$\%$ by mass envelope by the end of accretion from a gaseous disk. The disk gas abruptly dissipates after 3 Myr (vertical dotted line). A) Plot of interior mass fraction of H$_2$ with time compared with the concentration at the interface between the core and the envelope (dashed curve). B) Temperature variation with time (solid curve) compared with the consolute temperature of the binodal (dashed curve). C) Radial positions of the engulfment front (solid), the temperature and pressure conditions of the crest of the binodal (dashed), and the transit radius of the planet (dash-dot) with time. D) Variation in the mass fraction of the planet comprising the hydrogen-rich envelope with time. E) Mixing power required versus that available with time. F) Intrinsic luminosity carried through the envelope with time. The interval of disequilibrium hydrogen partitioning is delineated by the grey shading.}   
\label{fig:6M_50day}
\end{figure*}

Results are reported for a fiducial $6\,M_\oplus$ planet on a 50-day orbit. An H$_2$-rich envelope accretes onto the molten core, kept warm by the limits on radiative cooling imposed by the nascent envelope. With the parameters for binding energy retention and atmosphere accretion used here, the initial potential temperature is 3033 K and the peak primary atmosphere (H$_2$ + He) mass fraction is $\sim13\%$, accreted over the 3 Myr lifetime of the enveloping gaseous disk. As the planet accretes hydrogen, accretional heating raises the potential temperature to a maximum of $\sim8000$ K within the first $\sim0.5$ Myr (Equation \ref{eq:Tpot}). Thereafter, the planet cools.

\begin{figure*}[!t]
\centering
\includegraphics[width=0.85\textwidth]{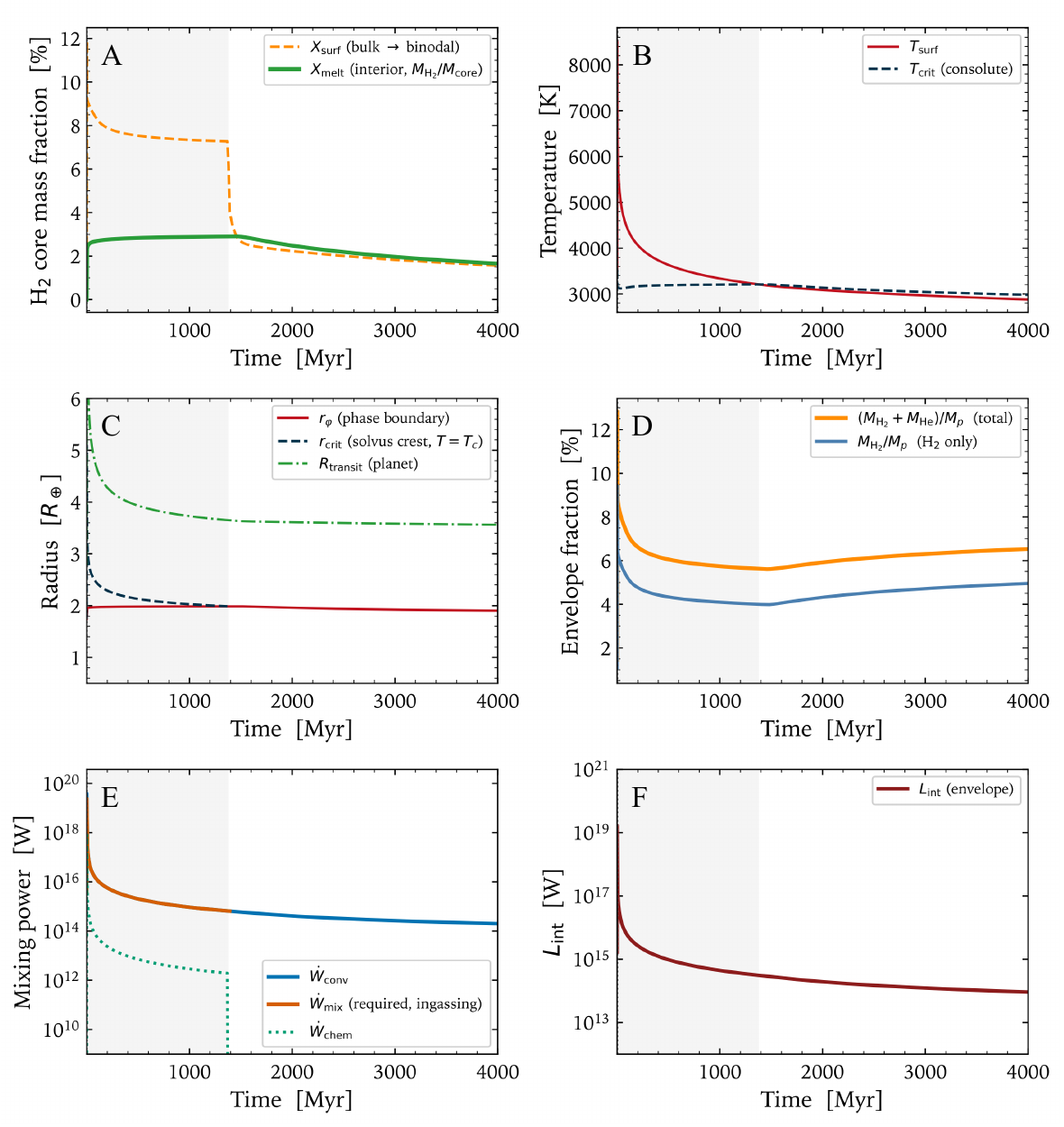}
\caption{Time evolution of the same 6$M_{_\oplus}$ planet in Figure \ref{fig:6M_50day} but shown with a linear time axis.  A) Plot of interior mass fraction of H$_2$ with time compared with the concentration at the interface between the core and the envelope (dashed curve). B) Temperature variation with time (solid curve) compared with the consolute temperature of the binodal (dashed curve). C) Radial positions of the engulfment front (solid), the temperature and pressure conditions of the crest of the binodal (dashed), and the transit radius of the planet (dash-dot) with time. D) Variation in the mass fraction of the planet composed of the envelope, and the hydrogen fraction of the envelope, with time.} E) Mixing power required versus that available with time. F) Intrinsic luminosity carried through the envelope with time.   
\label{fig:6M_50day_linear}
\end{figure*}

Figure \ref{fig:6M_50day} shows results for the fiducial $6\,M_\oplus$ planet. The mass fraction of hydrogen in the core rises mainly while disk gas is present (Figure \ref{fig:6M_50day}A; the dotted vertical line marks disk dispersal at $t = 3$ Myr), reaching $\sim 80\%$ of its eventual peak by that time. Immediately after dispersal, boil-off removes $\sim 26\%$ of the envelope H$_2$ within $\sim 0.3$ Myr before the cooling envelope contracts and the outflow returns to the luminosity-limited regime.  Ingassing then continues for another $\sim 1.4$ Gyr, more slowly, as the planet cools and the difference between the equilibrium and surface $x_{\rm H_2}$ values decreases (Figures \ref{fig:6M_50day}A, B). The interval of disequilibrium hydrogen partitioning is delineated by the grey shading in Figure \ref{fig:6M_50day}. Eventually, the outward moving engulfment front (solid curve, Figure \ref{fig:6M_50day}C) catches the inward moving binodal (dashed curve in Figure \ref{fig:6M_50day}C), at $T \simeq 3210$ K (Figure \ref{fig:6M_50day}B) and $P \simeq 7.2$ GPa, where the core hydrogen mass fraction reaches its peak of $\simeq 2.9\%$. The consolute $T$ in Figure \ref{fig:6M_50day}B is slightly greater than the binodal $T$ at the specific $x_{\rm sat}$ of the melt phase, reflecting the shape of the binodal in $T$ versus $x_{\rm H_2}$ space. During the subsequent cooling, a reversal occurs in which hydrogen shifts from the core back to the envelope in response to the decreasing binodal temperature. The shift is visible in Figure \ref{fig:6M_50day}D as an increase in the envelope mass fraction, from a minimum of $\sim 5.6\%$ to $\sim 6.5\%$ at 4 Gyr, by which time the core retains $\sim 1.7\%$ hydrogen. In this phase, the net flux of hydrogen across the interface is the small difference between the ingassing and degassing fluxes, and the slow, rotation-limited renewal of the magma ocean surface causes the melt to trail the binodal composition slightly, remaining supersaturated by $\lesssim 0.3\%$. The mixing of hydrogen into the core is energy limited for most of the accretion phase and throughout the post-accretion phase (Figure \ref{fig:6M_50day}E).Between $\sim0.2$ and 1 Myr, however, when accretional heating is strongest, the rate is instead limited by the convective overturn constrained by rotation. The intrinsic luminosity decreases by more than five orders of magnitude as the planet cools.

The same evolution on a linear time scale is shown in Figure \ref{fig:6M_50day_linear}, emphasizing the protracted changes over billions of years. Diffusion across the core--envelope interface keeps the melt close to the binodal at late times (Figure \ref{fig:6M_50day_linear}A), and the slow, modest decrease in the core hydrogen concentration with cooling is clearly seen. At its peak hydrogen concentration, the core retains $\sim 40\%$ of the hydrogen comprising the planet. The subsequent transfer of hydrogen from the core to the envelope raises the hydrogen envelope from a low of about $4\%$ of the planet mass at $\sim 1.5$ Gyr to $5\%$ at 4 Gyr (Figure \ref{fig:6M_50day_linear}D). The radius of the planet decreases by $\sim 2\%$ over this interval as the intrinsic luminosity declines (Figures \ref{fig:6M_50day_linear}C and F).

For sub-Neptune-like masses and orbital periods in general, the basic features of hydrogen engulfment and subsequent evolution are similar to those in Figures \ref{fig:6M_50day} and \ref{fig:6M_50day_linear}, with changes mainly in the timing of events. 
For example, for an orbital period of 20 days, the maximum concentration of H$_2$ in the supercritical melt is similar to the fiducial case presented here, all else equal, but occurs earlier at about 300 Myr post gaseous disk dispersal, and the retained hydrogen in the core after 4 Gyr is less, at about $0.5\%$ by mass, due to greater hydrogen loss resulting from proximity to the star. 
Engulfment is equally efficacious at large orbital radii, and is applicable to Neptune-like planets. 

\begin{figure}
\centering
\includegraphics[width=0.98\columnwidth]{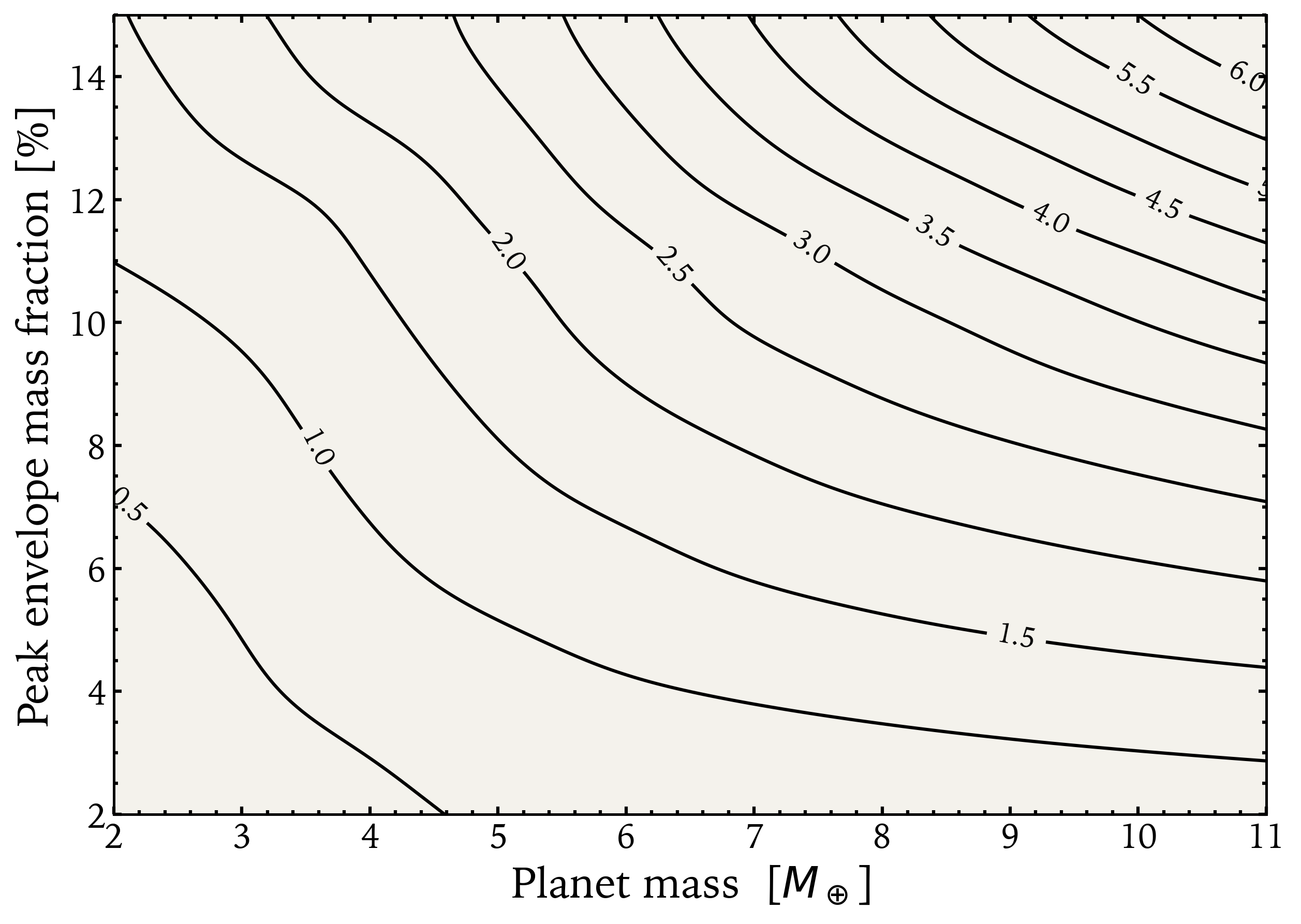}
\caption{Contours of model peak core hydrogen concentrations in weight percent as a function of planet mass and initial bulk H$_2$ concentration. The plot is for a fixed orbital period of 50 days and a rotation period of 17 hours.}
\label{fig:contours}
\end{figure}

The dependence of the maximal engulfed core-hydrogen mass fraction on planet mass and initial bulk hydrogen-rich atmosphere mass fraction is shown in Figure \ref{fig:contours}. The combination of planet mass and initial mass fractions of hydrogen acquired by the end of the accretion phase are the primary determining factors for the peak hydrogen contents of cores.  

\begin{figure*}
\centering
\includegraphics[width=0.99\textwidth]{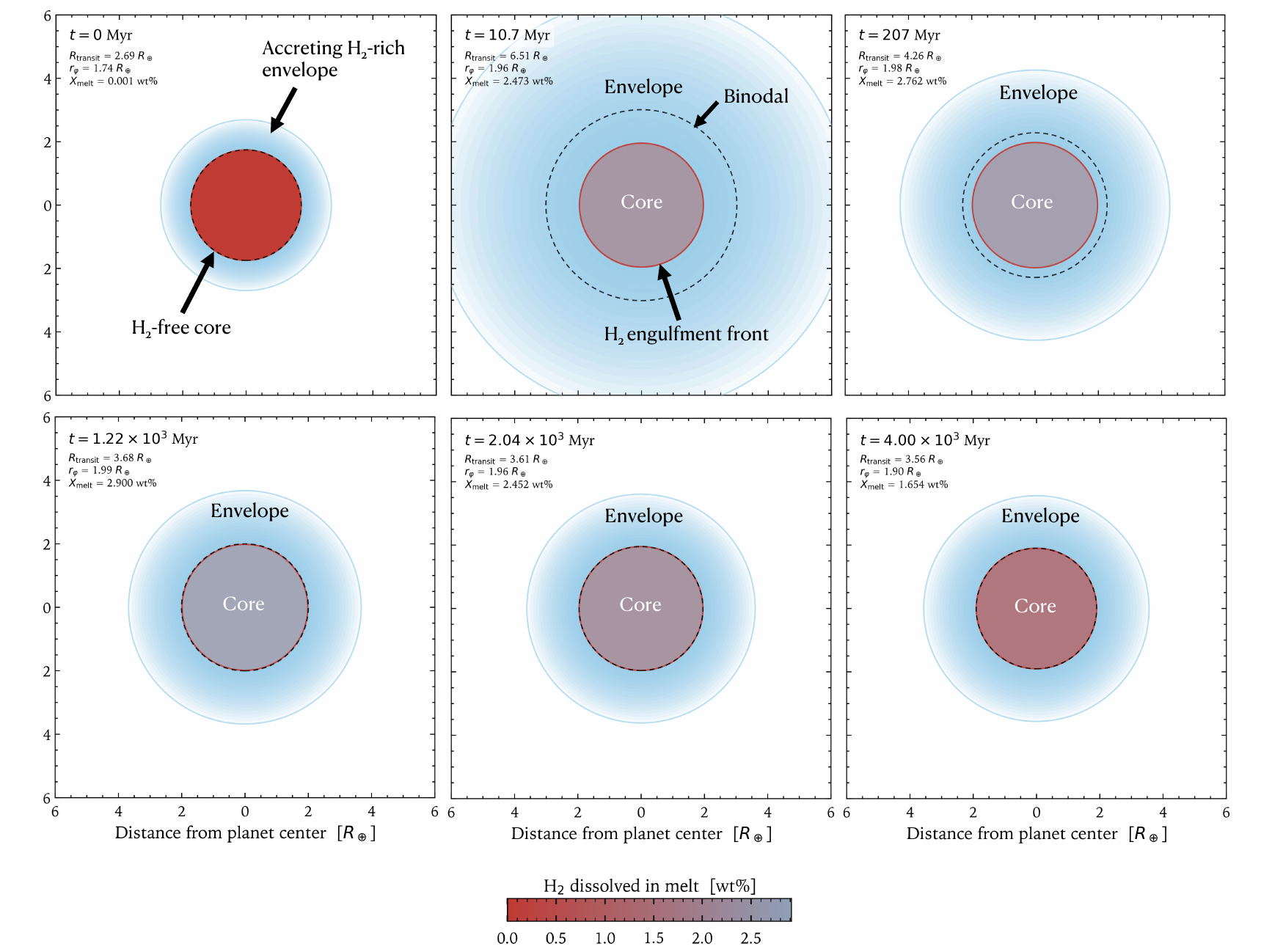}
\caption{Illustration of the evolution of sub-Neptune planets driven by hydrogen engulfment.  Each panel is a plot of a cross section through the planet at specified times for the model shown in Figures \ref{fig:6M_50day} and \ref{fig:6M_50day_linear}, with time moving from the upper left panel to the lower right panel. The colors indicate relative concentrations of hydrogen where blue is pure hydrogen.  In the post-accretion phase the engulfment front, the boundary between the supercritical magma ocean and overlying hydrogen-rich envelope (solid circle), slows its radial progression outward and the binodal, represented by the pressure-dependent consolute temperature surface (dashed circle), progresses inward, leading to the eventual coincidence of the binodal and the magma ocean surface. Thereafter, the binodal that now defines the interface between the core and envelope moves slowly inward during continued cooling.}
\label{fig:progress}
\end{figure*}
\section{Discussion}
\label{sec:discussion}

These results illustrate that molten cores and hydrogen envelopes should evolve towards chemical thermodynamic equilibration with respect to hydrogen concentrations given sufficient time.  The disequilibrium process of hydrogen engulfment drives hydrogen into the core, and full miscibility between silicate (and Fe, \citealt{Young2025_Differentiation}) and H$_2$ allows for large mass fractions of hydrogen in the planet interiors. In this scenario, no exogenous forcings to mix silicate melt and hydrogen are required.

Figure \ref{fig:progress} shows plots of the planet evolution shown in Figures \ref{fig:6M_50day} and \ref{fig:6M_50day_linear} in radius space in order to summarize the process of hydrogen engulfment resulting from diffusion coupled with convective surface renewal. Initially, a hydrogen-free core accretes hydrogen-rich gas ($t = 0$, Figure \ref{fig:progress}). As the envelope grows, the H$_2$-engulfment front that defines the interface between the core and envelope moves radially outward as hydrogen diffuses into the core and is redistributed by convective surface renewal ($t = 10.7$ Myr, Figure \ref{fig:progress}). Later, the outward progression of this front slows as more hydrogen builds up in the core and at the same time, the radial surface corresponding to the $T-P$ conditions for the silicate-hydrogen binodal (dashed circle in Figure \ref{fig:progress}) moves inward radially due to cooling ($t = 207$ Myr, Figure \ref{fig:progress}).  The two boundaries eventually coincide ($t = 1.2\times 10^3$ Myr, Figure \ref{fig:progress}), after which the binodal defining the core–envelope interface continues to migrate slowly inward ($t = 2.04\times 10^3$ to $4.00\times 10^3$ Myr, Figure \ref{fig:progress}). The latter stage in which the core-envelope interface moves inward with cooling has been modeled previously \citep{Rogers_2025_redefine, Young_2026} as a quasi-static sequence of states defined by the silicate-hydrogen binodal as the planet cools.

Comparing model results with the observed sub-Neptune population suggests that while the results in the upper right corner of Figure \ref{fig:contours} are feasible, they are not relevant to sub-Neptunes; radii for planets with masses of 10$M_{\oplus}$ and $10\%$ by mass hydrogen are significantly greater than the radius cliff of about 4$R_{\oplus}$. Other factors are at work that limit the total hydrogen in the case of the largest sub-Neptunes. The large mass fractions of hydrogen may be relevant to Neptune-like planets, however, if they harbor hydrogen-rich magma oceans beneath their hydrogen atmospheres. 

Our current understanding of the thermochemistry of silicate---iron---hydrogen mixing suggests that where hydrogen comprises greater than approximately $0.5\%$ by weight, and assuming solar-like fractions of iron and silicate, the core composed of these materials should be fully miscible at pressures greater than $\sim 10$ GPa \citep{Young2025_Differentiation}. Because magma ocean surface pressures are in the range of a few GPa, and increase with depth, virtually the entire phase space in Figure \ref{fig:contours} is indicative of deep interiors with no distinct iron cores. Subsequent degassing with cooling can in principle, under the right circumstances, lower hydrogen concentrations sufficiently to trigger silicate-iron differentiation and is worth further study.   

Limitations to ingassing by convective surface renewal could arise for at least two reasons based on these calculations.  Firstly, by assuming the efficiency of accretional heating is high ($\eta_{\rm eff} = 1$, a hot start), the peak temperature during accretion is ensured to be much greater than the equilibrium binodal temperature at a given pressure during the gaseous disk phase. Decreasing the efficiency of retention of heat during accretion to $0.5$ (half the accretion heat is lost to radiation) lowers the peak temperature and the maximal mass fraction of hydrogen in the core by about half that shown in Figure \ref{fig:6M_50day}, for example. For a truly cold start, with $\eta_{\rm eff} \sim 0.1$, relatively little hydrogen is engulfed by the core, e.g., $0.3\%$ by mass in the case shown in Figure \ref{fig:6M_50day}. In this case, the temperature at the core-envelope interface is roughly that of the binodal.  The mechanism proposed here thus requires that during accretion, the interface between the core and envelope must be hotter than the binodal if large mass fractions of hydrogen are to be imbibed by the core. 

The second mechanism that might limit, or at least reduce, hydrogen engulfment is formation of a durable, non-transient double-diffusive staircase or a stably stratified surface barrier. Further investigation into the longevity of such structures under these conditions is warranted. However, based on the discussion in \S \ref{sec:double}, structures such as these are generally thought to be transient, and so, where present, might slow the engulfment process but should not eliminate it. 

The results here are robust against uncertainties in hydrogen diffusivities. In principle, one might conclude that the precise numerical values for hydrogen diffusivities are crucial for the ingassing calculations, where the mass transfer coefficient for surface renewal varies with diffusivity according to $k_{\rm L} \propto\sqrt{D_{\rm H_2}}$. However, numerical results indicate that a 10-fold variation in hydrogen diffusivity produces shifts of \(\mathcal{O}(1\%)\) of the values for core H\(_2\) concentrations (e.g., a shift of $0.08$ in the $3.4\%$ by mass H$_2$ in Figure \ref{fig:6M_50day}A). This small dependency on values for $D_{\rm H_2}$ is the consequence of the system operating in the buoyancy-limited regime where $J_{\rm H}\rightarrow J_{\rm crit}$ and $J_{\rm crit}$ is independent of $D_{\rm H_2}$ (Equation \ref{eq:Jcrit}). Based on systematic differences between the hydrogen diffusivities in melt and those of melt structural components like Mg, Si and O that are $\sim 5\times$ smaller than those for H \citep{Karki_2010}, it seems likely that any systematic errors in the hydrogen diffusivities themselves are less than this difference.   

The protracted evolution seen in our models depends on the interior
remaining molten and convecting \citep{ginzburg2016a, Vazan2018}. Although most ingassing occurs during the gaseous disk phase,  long-lived convection sustains surface renewal well beyond disk
dispersal, drives the approach to the binodal equilibrium, and enables the late,
near-equilibrium back-transfer of hydrogen to the envelope. It also sets up the
inverse problem of when atmospheric escape strips the envelope. The models produced here include mass loss of the envelopes. However, the degree to which hydrogen in molten cores can be ``frozen" in with complete, or nearly complete, loss of hydrogen envelopes is a topic for a separate paper in preparation.  This process is of relevance to stripped cores that are thought to populate super-Earths.

\section{Conclusions}
\label{sec:conclusions}
 
 Convective surface renewal provides an intrinsic mechanism by which growing
planets with molten cores engulf hydrogen from their overlying H$_2$-rich
envelopes, requiring no exogenous forcing such as giant impacts. The engulfment
is governed by a Danckwerts mass-transfer coefficient in
which the renewal rate is determined by the planet's
convective overturn. Because the same top-down cooling that drives the overturn
continually sweeps the buoyant, hydrogen-bearing outer layer back into the interior, the
low-density boundary layer that would otherwise stall diffusive ingassing is
renewed rather than allowed to stagnate. Transfer of hydrogen from the envelope to the core is most effective during the gaseous disk phase where hydrogen is essentially in infinite supply, magma ocean surface temperatures are high, and the thermodynamic driving force for ingassing is strongest.

Hydrogen engulfment bears directly on the observable atmospheres and radii of
sub-Neptunes and super-Earths \citep{Young_2026}. In the case of sub-Neptunes, engulfment couples interior chemistry to the atmosphere in ways that are increasingly accessible to observation \citep{Nixon2025_TOI270d}. In the case of super-Earths, where they are stripped cores of sub-Neptunes, hydrogen stored in the molten interior lowers interior densities and inflates planetary radii.  The engulfment front is the
\emph{kinetic} route by which hydrogen-rich cores may form during accretion, and it
is complementary to the quasi-static, cooling-driven descent of the binodal that has
been modeled for the post-formation era \citep{Rogers_2025_redefine, Young_2026}. A
self-consistent description of sub-Neptune evolution therefore requires both the growth-phase
hydrogen engulfment front that moves outward as the core consumes hydrogen and the subsequent inward migration of the binodal during protracted cooling and liberation of hydrogen from the core.

The process of surface renewal may have applications to other aspects of magma ocean-envelope exchange, including silicate species and noble gases. Chemical forcings such as formation of water by hydrogen-silicate intramelt reactions can further aid in hydrogen engulfment and warrant  study. 

\begin{acknowledgments}
The author acknowledges thoughtful comments by an anonymous reviewer that led to improvements to the paper. Financial support provided by NASA grant 80NSSC24K0544. Portions of the Python model, analysis, and
figure-generation code were developed, debugged, and documented with the
assistance of Anthropic's Claude (Claude Opus 4.8, Anthropic, 2025--2026)
as an AI coding tool. The author designed the model, verified all code and
numerical results, and takes full responsibility for the content of this work.
\end{acknowledgments}

\bibliographystyle{aasjournal}
\bibliography{edy_references}

\end{document}